\documentclass[%
reprint, 
superscriptaddress, 
floatfix, longbibliography, 
amsfonts, amssymb, amsmath
]{revtex4-1}

\usepackage{graphicx}
\usepackage{bm}
\usepackage{dcolumn}
\usepackage{color}

\usepackage[
colorlinks=true, citecolor=blue, urlcolor=blue, linkcolor=blue,
setpagesize=false
]{hyperref}

\begin{document}
\title{Universal quantum criticality in the metal-insulator transition 
of two-dimensional interacting Dirac electrons
}

\author{Yuichi Otsuka}
\email{otsukay@riken.jp}
\affiliation{Computational Materials Science Research Team, 
RIKEN Advanced Institute for Computational Science (AICS), 
Kobe, Hyogo 650-0047, Japan}

\author{Seiji Yunoki}
\affiliation{Computational Materials Science Research Team, 
RIKEN Advanced Institute for Computational Science (AICS), 
Kobe, Hyogo 650-0047, Japan}
\affiliation{Computational Condensed Matter Physics Laboratory, 
RIKEN, Wako, Saitama 351-0198, Japan}
\affiliation{Computational Quantum Matter Research Team, 
RIKEN Center for Emergent Matter Science (CEMS), 
Wako, Saitama 351-0198, Japan}

\author{Sandro Sorella}
\affiliation{Computational Materials Science Research Team, 
RIKEN Advanced Institute for Computational Science (AICS), 
Kobe, Hyogo 650-0047, Japan}
\affiliation{SISSA -- International School for Advanced Studies, 
Via Bonomea 265, 34136 Trieste, Italy}
\affiliation{Democritos Simulation Center CNR--IOM Instituto Officina dei Materiali, 
Via Bonomea 265, 34136 Trieste, Italy}

\date{\today}

\begin{abstract}
 The metal-insulator transition has been a subject of intense research
 since Nevil Mott has first proposed that the metallic behavior of
 interacting electrons could turn to the insulating one as electron
 correlations increase. Here, we consider electrons with massless Dirac-like 
 dispersion in two spatial dimensions, described by the Hubbard models 
 on two geometrically different lattices,
 and perform numerically exact calculations on unprecedentedly large systems 
 that, combined with a careful finite size scaling analysis, 
 allow us to explore the quantum critical behavior 
 in the vicinity of the interaction-driven metal-insulator transition. 
 We find thereby that the transition is continuous and determine 
 the quantum criticality for the corresponding universality class, 
 which is described in the continuous limit by the Gross-Neveu model, 
 a model extensively studied in quantum field theory. 
 We furthermore 
 discuss a fluctuation-driven scenario for 
 the metal-insulator transition in the interacting Dirac electrons: 
 the metal-insulator transition is triggered only by the vanishing of the quasiparticle
 weight but not the  Dirac Fermi velocity, which instead remains 
 finite near the transition. 
 This important feature cannot be captured by  
 a simple mean-field or Gutzwiller-type approximate picture, 
 but is rather consistent with the low energy behavior of 
 the Gross-Neveu model.
\end{abstract}

\maketitle

\section{\label{sec:intro}Introduction}

The metal-insulator transition is one of the most fundamental 
and yet profound physical phenomena of quantum mechanics, and, 
in the absence of correlations, is described by the conventional 
band theory~\cite{Hoddeson_RMP1987,Imada_RMP1998}. 
A metal is such if electrons do not fill an integer number
of bands, otherwise insulating behavior settles  because an energy gap
is required to excite an electron from a fully occupied to an empty band.  
With this simple criterion, most insulating and metallic properties 
were successfully explained~\cite{Kittel_1967}.
However, it was soon realized by 
Mott~\cite{Mott_1949} in 1949 that the electron correlations could play a
major role in several materials, as they could become insulators 
even when the band theory predicts instead metals: 
these are the so called Mott insulators. 
Since then, many theoretical and numerical works have tried to shed
lights on this issue, but our understanding of 
interaction-driven metal-insulator transitions 
still remains rather controversial
because strongly correlated systems are hard to solve 
using both analytical and numerical methods, 
at least, when the spatial dimensionality is larger than one~\cite{Giamarchi_2004}
but smaller than infinity~\cite{Metzner_PRL1989}.

Gutzwiller has introduced in the middle 60's a correlated 
framework~\cite{Gutzwiller_1965}, 
that was later used to derive the properties of the metal-insulator
transition as a function of the correlation strength $U$.
This framework predicts for generic lattice models that for $U$ 
below the critical point $U_{\text{c}}$, 
the quasiparticle weight $Z$, which should be exactly
one in the non-interacting band theory, is strongly renormalized
by the correlation and vanishes as $Z \simeq (U_{\text{c}}-U)$. 
At the same time, the bandwidth $W$, renormalized by the electron correlations, 
reduces to zero at the transition in the same way as $Z$ vanishes.  
The prediction of the Brinkman-Rice approximation~\cite{Brinkman_PRB1970} 
has been later confirmed and further extended by the 
dynamical mean-field theory (DMFT)~\cite{Georges_RMP1996,Rosenberg},
an approach that is 
exact only in the limit of large spatial dimensions.

\begin{figure}[ht]
 \centering
 \includegraphics[width=0.40\textwidth]{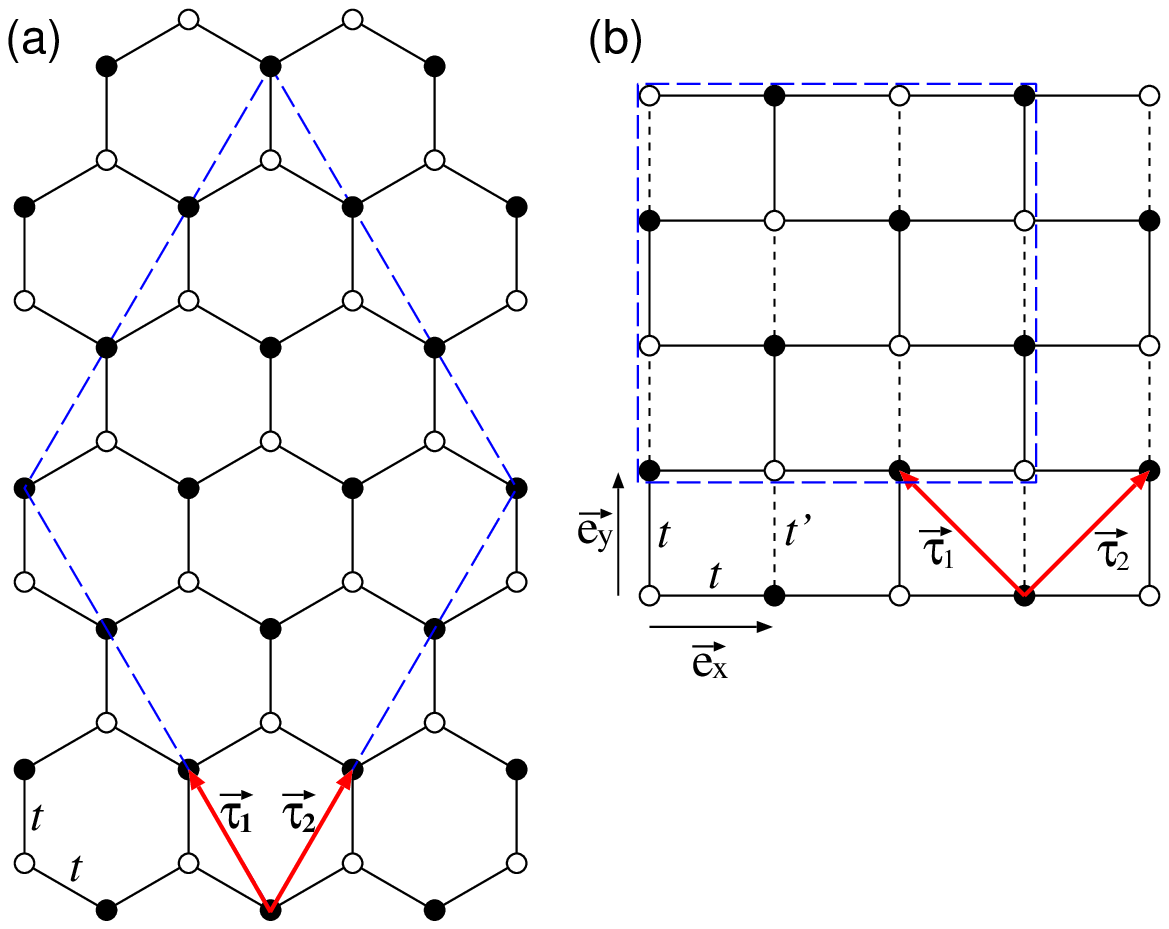}\\
 \includegraphics[width=0.23\textwidth]{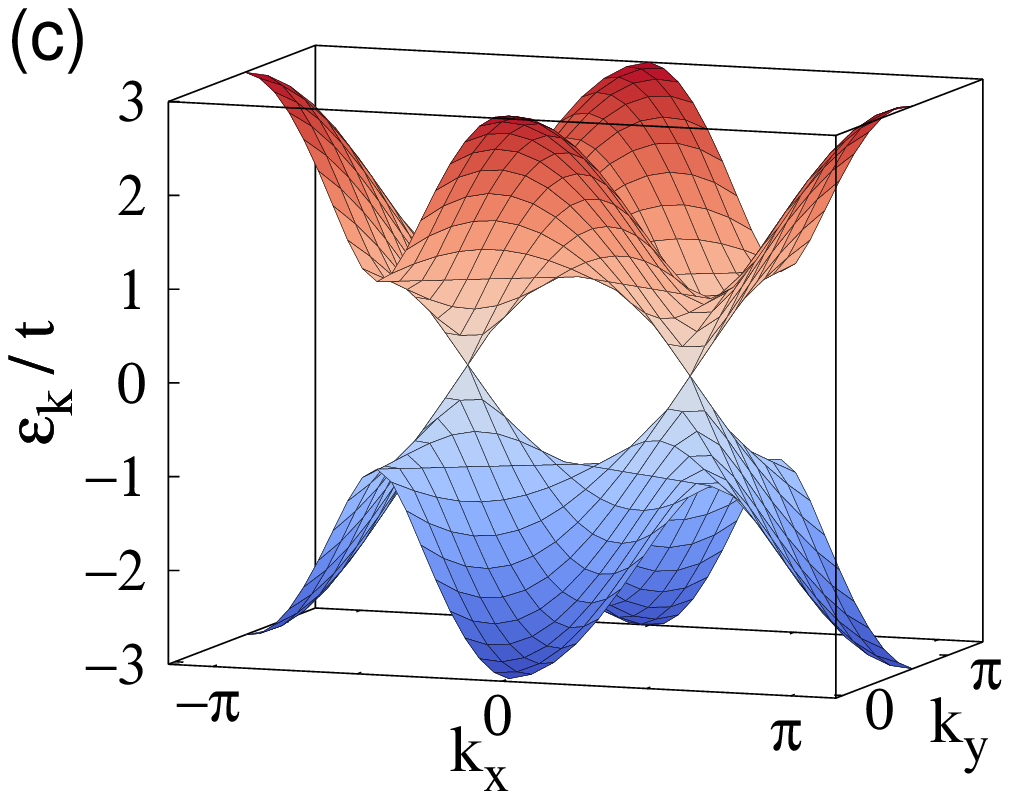}
 \includegraphics[width=0.23\textwidth]{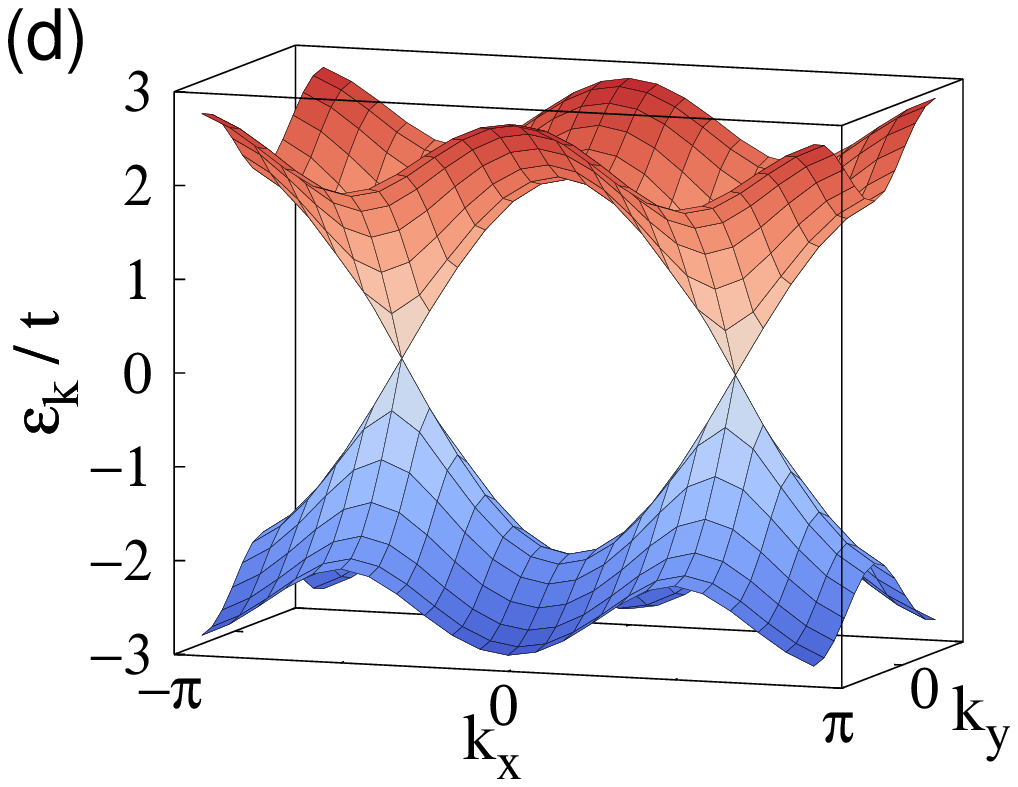}
 \caption{\label{fig:lattice}%
 Lattice structures
 for (a) the honeycomb lattice model 
 and (b) the $\pi$-flux model,
 where sites belonging to $A$ and $B$ sublattices are indicated 
 by solid and open circles. 
 The primitive translational vectors are denoted by 
 (a) ${\vec \tau}_1=(-\frac{\sqrt{3}}{2},\frac{3}{2})$ 
 and ${\vec \tau}_2=(\frac{\sqrt{3}}{2},\frac{3}{2})$, and 
 (b) ${\vec \tau}_1=(-1,1)$ and ${\vec \tau}_2=(1,1)$, 
 where the lattice constant between the nearest neighbor sites is set to be one. 
 The unit cells for both models thus contain two sites. 
 The nearest neighbor hopping parameters are indicated 
 by $t$ (solid lines) and $t'$ (dotted lines). $t'$ is set to be $-t$ for 
 the $\pi$-flux model in (b). 
 The cluster with $L=3$ ($L=4$) for the honeycomb ($\pi$-flux) model
 is indicated by blue dashed line. 
 The unit vector $\vec e_x$ ($\vec e_y$) along the $x$ ($y$) direction 
 is also indicated in (b). 
 The non-interacting energy dispersions $\varepsilon_{\bf{k}}$ 
 are shown 
 for (c) the honeycomb lattice model 
 and (d) the $\pi$-flux model. 
 The Fermi level is at $\varepsilon_{\bf{k}}=0$ for half-filling,
 and the Dirac points are located exactly at the Fermi
 level, where the valence and the conduction bands touch 
 with opposite chiralities. 
 Notice that in both models there are two distinct Dirac points at 
 (c) ${\bf{k}}=\frac{2\pi}{3}(\pm\frac{1}{\sqrt{3}},1)$ 
 and 
 (d) ${\bf{k}}=(\pm\frac{\pi}{2},0)$, 
 corresponding to two distinct valleys and 
 thus there are eight components of Dirac fermions in total 
 due to different chiral, spin, and valley degrees of freedom.  
 }
\end{figure}

Here, we focus on a specific realization of the metal-insulator transition 
in two-dimensional lattice models
which can be treated with a numerically exact method,
i.e., the Hubbard models defined on the honeycomb lattice 
and on the square lattice with $\pi$ flux penetrating each plaquette. 
These models are equipped with a free electron energy dispersion 
with nodal gapless points in the Brillouin zone and with linear dispersion 
(see Figs.~\ref{fig:lattice}),   
a very peculiar character of the so called ``massless Dirac electrons''. 
We set one electron per site, where the non-interacting band 
is half-filled 
and the Fermi surface is constituted by the Dirac points. 
Due to these gapless Dirac points, it is possible to have a non trivial 
metal-insulator transition at a finite value of the correlation strength
even in such bipartite lattices~\cite{square}.

Quite recently, a numerical simulation
of the Hubbard model on the honeycomb lattice has provided evidences
for a possible unconventional phase, 
a spin liquid phase with no classical order,
close to the metal-insulator transition
occurring at sufficiently large $U$~\cite{Meng_Nature2010}.
Although there is still activity~\cite{Chen_PRB2014},
the possibility of such an
intermediate phase between the semi-metal (SM) and 
the antiferromagnetic (AF) Mott insulator 
seems now rather unlikely, in view of 
the large scale simulations that we have reported recently~\cite{Sorella_SR2012}, 
clearly showing that the AF moment develops continuously from zero once 
the insulating phase is entered. 
Later studies have also confirmed the simplest scenario of a direct and continuous 
transition~\cite{Assaad_PRX2013,Hassan_PRL2013,Ixert_PRB2014,ParisenToldin_PRB2015}.

Similarly, a stable spin liquid in interacting Dirac
electrons represented by a different model has been also proposed 
in Ref.~\cite{Chang_PRL2012}.
Here, the  Hubbard model  on the square lattice is studied,
in which a flux  $\pi$ is added to each plaquette 
in order to obtain a massless Dirac dispersion in 
the non-interacting limit 
[Figs.~\ref{fig:lattice}(b) and \ref{fig:lattice}(d)] (referred to as 
the $\pi$-flux model hereafter). 
Based on an approximate numerical simulation with relatively small clusters, 
a spin liquid phase has been observed between the SM and the AF Mott insulator, 
which in this case only has a finite charge gap with
a vanishingly small spin gap~\cite{Chang_PRL2012}. 
This finding is significant also in the context of high-$T_{\text{c}}$ cuprate 
superconductors because the $\pi$-flux model is considered as one of the relevant 
models to understand the mechanism of 
superconductivity~\cite{AffleckMarston_PRB1988,AffleckAnderson_PRB1988}.
However, as in the case of the honeycomb lattice model,
this quantum disordered state has also been 
disputed~\cite{Otsuka_Proc2014,Ixert_PRB2014,ParisenToldin_PRB2015}.

Although it is obviously very important to 
search for spin liquid phases in ``realistic'' models, 
here we take a different perspective.
After several years of efforts on these strongly correlated systems, 
we feel that the time is mature to 
examine the quantum criticality in the metal-insulator transition of 
interacting electrons in two spatial dimensions, and in particular  the interacting 
Dirac electrons described by these two models where $U_{\text{c}}$ is 
finite and their ground state properties
can be explored by using 
an unbiased and formally exact numerical method.
This is precisely the main purpose of this paper.

Moreover, it has been recently shown that  
the interacting Dirac electrons on the honeycomb lattice can be mapped 
in the continuous limit onto a model well known in quantum field theory, i.e., 
the Gross-Neveu (GN) model~\cite{Gross_PRD1974} 
in the chiral Heisenberg universality class with $N=8$ 
fermion components~\cite{Herbut_PRL2006,Herbut_PRB2009a}. 
Since the $\pi$-flux model is expected to have a similar effective
theory in the continuous limit, 
it is reasonable to conjecture that 
the metal-insulator transition in these two specific lattice models 
belongs to the same universality class. 
In order to address this issue, 
here we perform large-scale quantum Monte Carlo (QMC) calculations
and evaluate the critical exponents with a high degree of accuracy.
This is indeed made possible because, 
with the help of the auxiliary field 
technique~\cite{Blankenbecler_PRD1981,Hirsch_PRB1985,White_PRB1989},
these fermionic models can be studied 
without the notorious ``sign problem''~\cite{Sorella_EPL1992,Otsuka_PRB2002}.
The careful finite size scaling analysis finds that 
the critical exponents for these two models are the same within statistical errors
and thus confirms the conjecture. 
Our results represents the first  accurate determination of 
the critical exponents for the 
GN model in the chiral Heisenberg universality class 
with 
$N=8$~\cite{Rosenstein_PhysLettB1993,Janssen_PRB2014,ParisenToldin_PRB2015}.

The other interesting issue to be addressed in this paper is to explore 
the quantum critical behavior in both metallic and insulating phases 
at the vicinity of the metal-insulator transition, in particular, 
the fate of the quasiparticle weight $Z$ and the Fermi velocity $v_{\text{F}}$
when approaching the critical point $U_{\text{c}}$ from the metallic side.
For electrons with the usual energy dispersions such as the one in the square lattice,
the Gutzwiller-type approximate description~\cite{Gutzwiller_1965,Brinkman_PRB1970}
and the simple DMFT approach~\cite{Georges_RMP1996,Rosenberg} 
predict that $Z$ and $v_{\text{F}}$ are both renormalized by the interaction,
and vanish at $U_{\text{c}}$. 
This scenario is valid for any lattice model and in any dimensionality 
within the Gutzwiller approximation since, within this method, 
the free electron dispersion is simply renormalized by a Gutzwiller factor $Z$ 
that vanishes at the transition.  
Analogously, the same scenario holds within the single-site DMFT~\cite{Tran_PRB2009}
because, once the self-energy is assumed to be momentum-independent, 
the free electron dispersion can be renormalized 
only through the quasiparticle weight $Z$~\cite{Georges_RMP1996}.
Instead, our unbiased and numerically exact calculations support the  
qualitative prediction based on the renormalization group (RG) analysis
for the GN model~\cite{Herbut_PRB2009b} 
and the recent numerical results for the honeycomb lattice model 
obtained by advanced quantum cluster methods~\cite{Wu_PRB2010,Wu_PRB2014}:
with increasing the correlation strength, $Z$ vanishes at the transition, 
while the Fermi velocity $v_{\text{F}}$ remains finite.

Our large-scale QMC calculations also provide a firm numerical evidence for the absence of 
a spin liquid phase in between the SM and the AF insulator for the $\pi$-flux model, 
thus ruling out the possibility of a spin liquid phase reported previously in Ref.~\cite{Chang_PRL2012}. 
This is very similar to the case for the honeycomb lattice model, 
where the originally proposed spin liquid phase~\cite{Meng_Nature2010} is 
turned out to be rather implausible
after our large-scale calculations~\cite{Sorella_SR2012}. 
The metal-insulator transitions in both models are rather direct and continuous, 
and can be characterized by the quantum critical behavior of 
the quasiparticle weight in the metallic phase 
and the antiferromagnetic order parameter in the insulating phase. 
These results therefore suggest that the electron correlation alone is 
not enough but other factors such as geometrical 
frustration are required for a magnetically disordered spin liquid state~\cite{Balents_2010}.

The rest of this paper is organized as follows.
The definition of the two models 
and a brief description of the QMC method employed 
are given in Sec.~\ref{sec:model}. 
The ground state phase diagrams are first obtained in Sec.~\ref{sec:poorman} 
by a rather conventional way of extrapolating order parameters to 
the thermodynamics limit. 
Section~\ref{sec:fss} is devoted to more detailed analysis 
to determine the critical exponents with high accuracy. 
The fate of the Fermi velocity is investigated in Sec.~\ref{sec:vf}. 
Finally, the results are discussed in the context of the GN model, 
followed by an outlook and conclusions, in Sec.~\ref{sec:conclusion}. 
The energy resolved momentum distribution function is described in 
Appendix~\ref{appsec:nk_def} and the leading correction to 
the scaling analysis is discussed in Appendix~\ref{appsec:correction}.

\section{\label{sec:model}Models and Method}

We consider two variants of the Hubbard models in two spatial dimensions, 
whose low-lying energy states are described by 
the interacting Dirac fermions with spin 1/2 degree of freedom at half-filling.
The Hamiltonian in standard notations reads
\begin{equation}
 \hat H= -\sum\limits_{\langle i,j\rangle}\sum_{s=\uparrow,\downarrow} 
   t_{ij} c^{\dag}_{is} c_{js} 
 + U\sum_{i} n_{i\uparrow} n_{i\downarrow},
\label{eq:hamiltonian}
\end{equation}
where $c_{is}^\dag$ is the creation operator of electron at site $i$ and 
spin $s\,(=\uparrow,\downarrow)$, $n_{is}=c_{is}^\dag c_{is}$, and 
the sum $\langle i,j\rangle$ runs over all pairs of nearest neighbor sites $i$ and $j$.
The first model is defined on the honeycomb lattice with the uniform hopping 
$t_{ij}=t$ [see Fig.~\ref{fig:lattice}(a)]. 
The second one is on the square lattice with a flux of $\pi$ penetrating 
on each square plaquette, represented, with an appropriate gauge transformation, 
by $t_{i,i+\vec{e}_x}=t$ and $t_{i,i+\vec e_y}=(-1)^{i_x+i_y}t$, 
where the position of site $i$ is given as 
$i_{x} \vec{e}_{x} + i_{y} \vec{e}_{y}$ and $\vec e_x$ ($\vec e_y$) denotes 
the unit vector along the $x$ ($y$) direction [see Fig.~\ref{fig:lattice}(b)].

The clusters considered here consist of 
$ (L \vec \tau_1, L \vec \tau_2)$ with $N_{\text{s}}=2L^2$ sites for the honeycomb lattice model 
and $(L\vec e_x,L\vec e_y)$ with $N_{\text{s}}=L^2$ sites for the $\pi$-flux model, 
as indicated in Figs.~\ref{fig:lattice}(a) and \ref{fig:lattice}(b), respectively, 
with periodic boundary conditions.
The number of electrons are set to be equal to the number of sites in both models. 
In order to include the Dirac points among the allowed momenta 
in the non-interacting energy dispersions, 
$L$ is chosen to be a multiple of three (four) 
for the honeycomb lattice ($\pi$-flux) model. 
The smallest clusters are indicated by dashed lines 
in Figs.~\ref{fig:lattice}(a) and \ref{fig:lattice}(b). 
The largest clusters considered here are 
$N_{\text{s}}=2,592$ sites for the honeycomb lattice model 
and $N_{\text{s}}=1,600$ for the $\pi$-flux model.

Although the two models are quite different, 
they are both characterized by the non-interacting energy dispersions 
$\varepsilon_{\bm{k}}$
with two gapless Dirac cones, as shown in Figs.~\ref{fig:lattice}(c) and \ref{fig:lattice}(d), 
leading to a semi-metallic behavior at half-filling and for small coupling 
$U/t$~\cite{Sorella_EPL1992,Otsuka_PRB2002}. 
The effective low-energy Hamiltonian $\hat H_{\text{eff}}^{0}$ in the non-interacting limit 
at the vicinity of the Dirac points for spin $s$ is described as 
\begin{equation}
 \hat H_{\text{eff}}^{0}=v_{\text{F}}^{0}(\pm\delta k_x\sigma_x+\delta k_y\sigma_y),
\label{eq:heff}
\end{equation} 
where $\delta{\bm{k}}=(\delta k_x,\delta k_y)$ is 
the momentum measured from the Dirac point, 
$v_{\text{F}}^{0}=3t/2\,(2t)$ 
is the Dirac Fermi velocity in the non-interacting limit 
for the honeycomb lattice ($\pi$-flux) model, 
and $\vec\sigma=(\sigma_x,\sigma_y,\sigma_z)$ are the Pauli matrices 
acting on the two different sublattices.  
As shown below, both models display the metal-insulator transitions at finite critical
values $U_{\text{c}}/t$ from the non-magnetic SM to the AF long-range ordered 
insulating phase, 
in good agreement with previous numerical studies~\cite{Sorella_EPL1992,Otsuka_PRB2002}. 
It should also be noted that in the simplest mean-field picture the insulating phase emerges 
because the mass term proportional to $\sigma_z$ is introduced Eq.~(\ref{eq:heff}) 
when an AF order sets in. 
Therefore, there is no unit cell doubling in the AF insulating 
phase for both models~\cite{note_MI}.

We employ the auxiliary field QMC
method~\cite{Blankenbecler_PRD1981,Hirsch_PRB1985,White_PRB1989}
to investigate the ground state properties of these two models.
The expectation value of a physical observable $\hat{\mathcal{O}}$ 
over the ground state $|\Psi_0\rangle$ of $\hat H$
is obtained by projecting out trial wave functions to the ground state, i.e., 
\begin{equation}
\langle \hat{\mathcal O} \rangle =  
\langle \Psi_0| \hat{\mathcal O} |\Psi_0 \rangle
= \lim_{\tau\to\infty} O(\tau), 
\end{equation}
where 
\begin{equation}
O(\tau)
= 
\frac{ 
\langle \psi_{\text{L}} |
\ e^{- \frac{\tau}{2} \hat{H} } 
\ \hat{O} 
\ e^{- \frac{\tau}{2} \hat{H}}
\ | \psi_{\text{R}} \rangle 
}
{ 
\langle \psi_{\text{L}}|
\ e^{-\tau \hat{H}}
\ | \psi_{\text{R}} \rangle 
}
\label{eq:projection}
\end{equation}
and $| \psi_{\text{L}} \rangle$ ($| \psi_{\text{R}} \rangle $) is 
the left (right) trial wave function, 
chosen to have finite overlap with the exact ground state.
We choose $| \psi_{\text{L}} \rangle$ as a mean-field wave function 
of Eq.~(\ref{eq:hamiltonian}) with an AF order parameter in the $x$
direction, while a Slater determinant of the non-interacting Hamiltonian 
is used for $| \psi_{\text{R}} \rangle $ to which a tiny perturbation
term is added to remove the degeneracy at the two Dirac points.
These choices for the trial wave functions have been shown to yield
a particularly fast convergence in the imaginary time 
projection onto the ground state~\cite{Sorella_SR2012}.

\begin{figure}[htbp]
 \centering
 \includegraphics[width=0.45\textwidth]{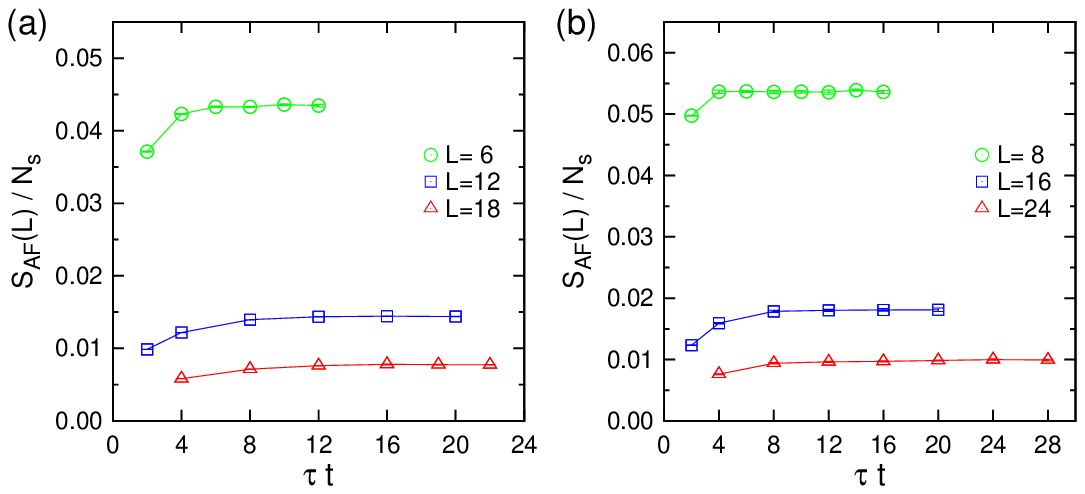}\\
 \includegraphics[width=0.45\textwidth]{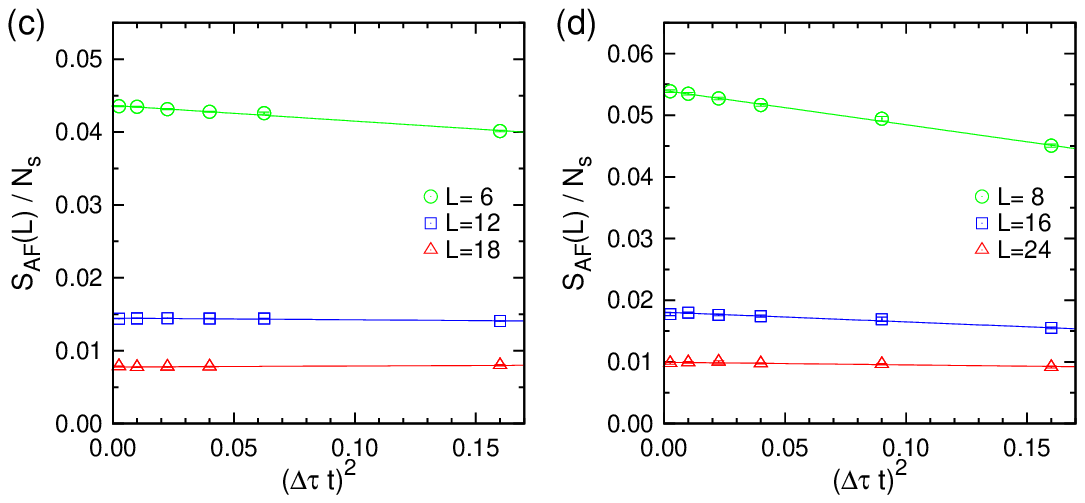}
 \caption{\label{fig:check}%
 Upper panels:
 Convergence of the spin structure factor $S_{\text{AF}}(L)$, 
 defined in Eq.~(\ref{eq:s_af}),
 with respect to the projection time $\tau$
 for (a) the honeycomb lattice model 
 and (b) the $\pi$-flux model.
 Lower panels:
 Extrapolation of $S_{\text{AF}}(L)$ to $\Delta \tau$=0
 for (c) the honeycomb lattice model 
 and (d) the $\pi$-flux model.
 The values shown in (a) and (c) for the honeycomb lattice model 
 are partially taken from  Ref.~\cite{Sorella_SR2012}. 
 We set $U/t=4$ $(5.8)$ for the honeycomb lattice ($\pi$-flux) model. 
 The cluster sizes $L$ used are indicated in the figures. 
 The statistical errors are smaller than the size of symbols. 
 Straight lines in (c) and (d) are least-square fits to the data 
 with linear functions of $(\Delta\tau t)^2$ for different $L$, 
 whereas lines in (a) and (b) are guides to the eye. 
 }
\end{figure}

The imaginary time evolution operator $e^{-\tau\hat{H}}$ with 
the projection time $\tau$ is divided into $N_\tau$ pieces, i.e., 
$e^{-\tau\hat H}=\left( e^{-\Delta\tau\hat H}  \right)^{N_\tau}$,
where $\tau = \Delta\tau N_{\tau}$ 
and $N_{\tau}$ is the Trotter number (integer). 
By setting $\Delta\tau t\ll 1$, we can use the Suzuki-Trotter 
decomposition~\cite{Trotter_1959,Suzuki_1976}, 
$e^{-\Delta\tau\hat H}=e^{-\frac{1}{2}\Delta\tau\hat H_0}e^{-\Delta\tau\hat H_{\text{I}}}e^{-\frac{1}{2}\Delta\tau\hat H_0}+{\cal O}\left(\Delta\tau^3\right)$, 
where $\hat H_0$ is the hopping term 
and $\hat H_{\text{I}}$ is the interacting term 
of the Hubbard model $\hat H$ in Eq.~(\ref{eq:hamiltonian}). 
Notice that the systematic error introduced in this decomposition 
is ${\cal O}\left(\Delta\tau^3\right)$. 
The discrete Hubbard-Stratonovich transformation is applied 
to $^{-\Delta\tau\hat H_{\text{I}}}$, 
which introduces an auxiliary Ising field at each site 
as well as at each imaginary time slice~\cite{Hirsch_1984}. 
As shown in Fig.~\ref{fig:check},
we have confirmed that the systematic errors 
due to finite $\tau$ and $\Delta\tau$ 
are sufficiently small, 
compared to the statistical errors in Monte Carlo importance sampling, 
when we choose $\tau=L+4$ and $\Delta\tau t =0.1$. 
More technical details are found in our previous report~\cite{Sorella_SR2012}.

\section{\label{sec:poorman}Ground state phase diagram}

In this section, we shall focus on the continuous nature of 
the quantum phase transition between the SM and the AF insulator. 
For this purpose, we calculate two fundamental quantities, 
the staggered magnetization and the quasiparticle weight, 
which characterize two different aspects across the transition.
The former quantity reveals the magnetic transition to the AF state
and the latter one directly captures the metal-insulator transition.
These transitions are expected to occur at the same critical $U_{\text{c}}$, 
unless there is an intermediate phase such as the spin
liquid phases~\cite{Meng_Nature2010,Chen_PRB2014,Chang_PRL2012,Sorella_SR2012,Hassan_PRL2013,Assaad_PRX2013}.
Here in this section we take a conventional and straightforward way, 
i.e., by first calculating the staggered magnetization 
and the quasiparticle weight on different finite clusters 
and then extrapolating them in the thermodynamic limit, 
to obtain the ground state phase diagram as a function of $U/t$.

\subsection{\label{subsec:poorman_m}staggered magnetization}

The staggered magnetization on each finite cluster 
with a linear dimension $L$, expressed as $m_{\text{s}}(L)$, is 
calculated from the spin structure factor $S_{\text{AF}}(L)$, i.e., 
\begin{equation}
 m_{\text{s}}(L)  = \sqrt{\frac{S_{\text{AF}}(L)}{N_{\text{s}}}}, 
 \label{eq:m_s}
\end{equation}
where
\begin{equation}\label{eq:s_af}
 S_{\text{AF}}(L) = \frac{1}{N_{\text{s}}}
 \left\langle
 \left(
 \sum_{i\in A} 
 \vec S_i
 -
 \sum_{i\in B} 
 \vec S_i
 \right)^{2}
 \right\rangle,
\end{equation}
$\vec S_i=\frac{1}{2}\sum_{s,s'}c^\dag_{is}(\vec\sigma)_{ss'}c_{is'}$ is 
the spin operator at site $i$, 
and the sum $i\in A(B)$ in Eq.~(\ref{eq:s_af}) runs over sites belonging to $A$ $(B)$ sublattices 
[see Figs.~\ref{fig:lattice}(a) and \ref{fig:lattice}(b)]. 
The AF order parameter $m_{\text{s}}$ is determined 
by extrapolating $m_{\text{s}}(L)$ in the thermodynamic limit, i.e., 
\begin{equation}
m_{\text{s}} = \lim_{L\to\infty} m_{\text{s}}(L). 
\end{equation}

\begin{figure}[htbp]
 \centering
 \includegraphics[width=0.48\textwidth]{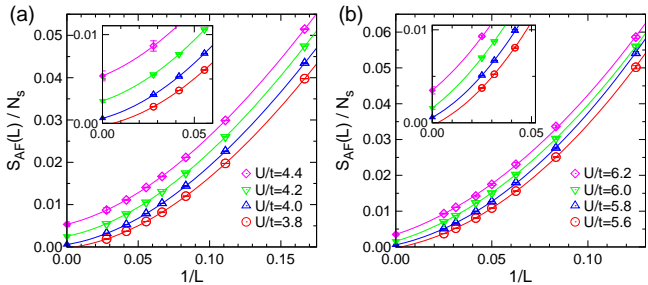}
 \caption{\label{fig:m2vsL}%
 Extrapolation of the spin structure factor $S_{\text{AF}}(L)$ to the thermodynamic limit
 for (a) the honeycomb lattice model 
 and (b) the $\pi$-flux model 
 with different $U/t$ values indicated in the figures.
 Solid curves are least-square fits of the data by cubic (quadratic)
 polynomials in $1/L$ for the honeycomb lattice ($\pi$-flux) model.
 Insets show enlarged plots for large $L$. The extrapolated values to 
 the thermodynamic limit are also indicated at $1/L=0$. 
 }
\end{figure}

\begin{figure}[ht]
 \centering
 \includegraphics[width=0.48\textwidth]{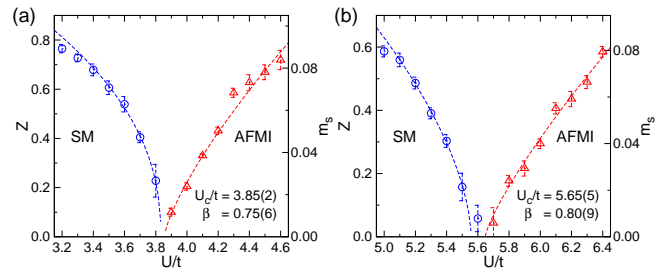}
 \caption{\label{fig:Uc}%
 The ground state phase diagrams 
 for (a) the honeycomb lattice model 
 and (b) the $\pi$-flux model. 
 Open triangles and open circles represent 
 the AF order parameter $m_{\text{s}}$ and 
 the quasiparticle weight $Z$, respectively.
 The critical $U_{\text{c}}$ estimated from $m_{\text{s}}$, assuming a form 
 of $m_{\text{s}}\sim (U-U_{\text{c}})^\beta$ (red dashed lines), 
 is indicated in the
 figures along with the critical exponent $\beta$.
 The chi-square values divided by degrees of freedom
 for this estimation are 0.88 for (a) and 1.05 for (b).
 Blue dashed line for $Z$ is a guide to the eye. 
 SM and AFMI stand for 
 semi-metal and antiferromagnetic insulator, respectively. 
 Most of values for $m_{\text{s}}$ 
 shown in (a) for the honeycomb lattice
 model are taken from  Ref.~\cite{Sorella_SR2012}.
 }
\end{figure}

At a fixed value of $U/t$, the staggered magnetizations $m_{\text{s}}(L)$ are calculated 
on clusters with $L=6$,  9, 12, 15, 18, 24, and 36 for the honeycomb lattice model, 
and with $L=8$, 12, 15, 18, 24, 32, and 40 for 
the $\pi$-flux model, and are extrapolated to the thermodynamic limit 
using polynomial functions in $1/L$. 
The typical results are shown in Fig.~\ref{fig:m2vsL}. 
It is observed 
that such a simple functional form represents the data rather well 
and thus we can estimate the AF order parameter $m_{\text{s}}$ reasonably accurately. 
The results for the extrapolated AF order parameter $m_{\text{s}}$ 
are summarized in Fig.~\ref{fig:Uc}.

The critical points $U_{\text{c}}$, 
above which the AF order parameter $m_{\text{s}}$ is finite,
and the critical exponents $\beta$ are estimated 
by assuming a form of the AF order parameter as a function of $U$ as 
\begin{equation}
m_{\text{s}} \sim (U- U_{\text{c}})^{\beta}.
\end{equation}
The estimated $U_{\text{c}}/t$ is $3.85\pm0.02$ for the honeycomb lattice 
model and $5.65\pm0.05$ for the $\pi$-flux model, as indicated in Fig.~\ref{fig:Uc}.  
Notice that $U_{\text{c}}$ for the $\pi$-flux model is larger 
than that for the honeycomb lattice model. 
This is easily understood because $v_{\text{F}}^0$ for the former model 
is larger than that for the latter model 
[see Eq.~(\ref{eq:heff}) and also Figs.~\ref{fig:lattice}(c) and \ref{fig:lattice}(d)] 
and thus a larger $U$ is required to induce the AF order. 
Although $U_{\text{c}}/t$ is different for these two models, 
our calculations find that the critical exponents $\beta$ for the two models 
are the same within statistical errors, 
i.e., $\beta=0.75\pm0.06$ 
for the honeycomb lattice model 
and $0.80\pm0.09$ for the $\pi$-flux model, 
as indicated in Fig.~\ref{fig:Uc}, which will be confirmed in more details 
by a careful and accurate finite-size scaling analysis in Sec.~\ref{sec:fss}.

\subsection{\label{subsec:poorman_z}quasiparticle weight}

\begin{figure}[htbp]
 \centering
 \includegraphics[width=0.48\textwidth]{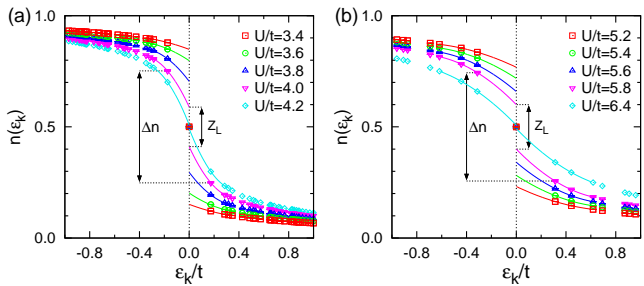}
 \caption{\label{fig:nk}%
 Energy resolved momentum distribution function 
 $n(\varepsilon_{\bm{k}})$ near the Fermi level, 
 indicated by a vertical dotted line at $\varepsilon_{\bm{k}}=0$, 
 for (a) the honeycomb lattice model ($L=36$) 
 and (b) the $\pi$-flux model ($L=40$) 
 with different values of $U/t$ indicated in the figures. 
 Solid curves are least-square fits of the three data points 
 closest to the Fermi level with $\varepsilon_{\bm{k}}<0$ and 
 $\varepsilon_{\bm{k}}>0$ using quadratic polynomials of $\varepsilon_{\bm{k}}$. 
 Notice that, due to the particle-hole symmetry, 
 $n(\varepsilon_{\bm{k}}>0) = 1-n(\varepsilon_{\bm{k}}<0)$ 
 and thus $n(\varepsilon_{\bm{k}}=0)=1/2$. 
 $Z_L$ and $\Delta n(u,L)$ are indicated in 
 (a) for $U/t=4$ and in (b) for $U/t=5.8$. 
 }
\end{figure}

Next, we study the metal-insulator transition 
by considering the momentum distribution function, i.e., 
the ground state occupation of the one electron states 
labeled by momentum $\bm{k}$, spin $s$, 
and non-interacting energy $\varepsilon_{\bm{k}}$.
We calculate this occupation number $n(\varepsilon_{\bm{k}})$ [defined in Eq.~(\ref{eq:nk})] 
as a function of $\varepsilon_{\bm{k}}$, 
where we average over equivalent momenta with the same energy 
(see Appendix~\ref{appsec:nk_def} for the details).
Typical results of the ``energy resolved'' momentum distribution function 
$n(\varepsilon_{\bm{k}})$ are shown in Fig.~\ref{fig:nk} for both models.
For $U<U_{\text{c}}$, a jump in the momentum distribution function 
$n(\varepsilon_{\bm{k}})$ occurs for $\varepsilon_{\bm{k}} \to 0$ 
when ${\bm{k}}$ approaches the Dirac points. 
The singularity in $n(\varepsilon_{\bm{k}})$ implies a long-distance 
power-law behavior in the density matrix 
$\langle c_{is}^\dag c_{js}\rangle$ in real space, 
which is the fingerprint of a metal.
From general grounds~\cite{singularity}, 
by applying the  well known ``Migdal theorem''~\cite{Migdal_JETP_1957}, 
the quasiparticle weight $Z$ can be related to the 
jump in the momentum distribution function. 
Therefore, we can have direct access to $Z$ 
in the thermodynamic limit. 
However, the finite size effects are rather significant 
and need to be carefully controlled 
in order to reach definite conclusions 
based on the available finite size calculations.

It should also be noticed in Fig.~\ref{fig:nk} that 
the energy resolved momentum distribution function $n(\varepsilon_{\bm{k}})$ 
becomes smooth without a visible discontinuity at the Fermi level for large $U/t$. 
This is interpreted as an exponential decay of the density matrix at large distance, 
because the density matrix is just the Fourier transform of the momentum distribution function. 
This clearly indicates the presence of a gap in the charge sector. 

We find that the following procedure works for estimating $Z$ 
in the thermodynamic limit. 
Since the ``energy resolved'' momentum distribution function 
$n(\varepsilon_{\bm{k}})$ is smooth near $\varepsilon_{\bm{k}} = 0$, 
we analyze $n(\varepsilon_{\bm{k}})$ calculated
on different finite clusters 
by extrapolating it to $\varepsilon_{\bm{k}} = 0$ with a polynomial function. 
More specifically,
the quasiparticle weight on each finite cluster, $Z_{L}$, is
defined as the jump in $n(\varepsilon_{\bm{k}})$,
which is evaluated by extrapolating 
to the Fermi level ($\varepsilon_{\bm{k}}=0$)
the closest three data points of $n(\varepsilon_{\bm{k}})$ for 
$\varepsilon_{\bm{k}}<0$ or $\varepsilon_{\bm{k}}>0$ 
with a quadratic polynomial, as shown in Fig.~\ref{fig:nk}.
Notice that due to the particle-hole symmetry
the two extrapolations using the data points for $\varepsilon_{\bm{k}}<0$ and 
for $\varepsilon_{\bm{k}}>0$ are related 
and in practice only one of the two is necessary.

\begin{figure}[htbp]
 \centering
 \includegraphics[width=0.48\textwidth]{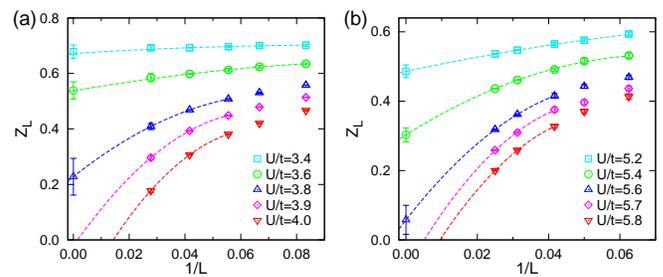}
 \caption{\label{fig:ZvsL}%
 Extrapolation of the quasiparticle weight $Z_L$ to the thermodynamic limit
 for (a) the honeycomb lattice model 
 and (b) the $\pi$-flux model 
 with different values of $U/t$ indicated in the figures. 
 Dashed curves are least-square fits for the largest sizes
 with quadratic polynomials of $1/L$, 
 which are extrapolated to the thermodynamic limit to estimate the quasiparticle 
 weight $Z$ indicated at $1/L=0$.
 }
\end{figure}

The extrapolated quantity $Z_L$ for finite $L$ 
is certainly much closer to the value in the thermodynamic limit 
and quite generally converges smoothly to the quasiparticle weight 
\begin{equation} 
Z= \lim \limits_{L \to \infty} Z_L
\end{equation} 
with a quadratic polynomial function of $1/L$, 
as shown in Fig.~\ref{fig:ZvsL}. 
However, this extrapolation is valid only in the metallic regime, where
the assumed polynomial convergence in $1/L$ is justified. 
Indeed, we find in Fig.~\ref{fig:ZvsL} that, for larger $U/t$,
$Z_L$ is extrapolated to a negative value, 
instead of being positive. 
This is clearly an inconsistent extrapolation 
because $Z_L$ in the insulating phase is expected to converge exponentially. 
Nevertheless, these inconsistent extrapolations are very useful 
as they allow us to identify the extension of the metallic region 
and determine the critical 
$U_{\text{c}}^{\text{MI}}$ for the metal-insulator transition.
It is indeed estimated in Fig.~\ref{fig:ZvsL} that 
$U_{\text{c}}^{\text{MI}}/t\sim3.9$ for the honeycomb lattice model 
and $U_{\text{c}}^{\text{MI}}/t\sim5.6$ for the $\pi$-flux model, 
which are in excellent agreement with the critical $U_{\text{c}}$ 
for the AF order (see Fig.~\ref{fig:Uc}).

The results for the quasiparticle weight $Z$ in the thermodynamic limit
as a function of $U/t$ are summarized in Fig.~\ref{fig:Uc} for both models.
The obtained $Z$ for $U<U_{\text{c}}^{\text{MI}}$ appears well behaved
and can provide the critical $U_{\text{c}}^{\text{MI}}$ 
by assuming that 
\begin{equation}
Z \sim (U_{\text{c}}^{\text{MI}}-U)^{\eta_{Z}} 
\label{eq:z_eta}
\end{equation} 
for $U<U_{\text{c}}^{\text{MI}}$ close to $U_{\text{c}}^{\text{MI}}$. 
We find that 
$U_{\text{c}}^{\text{MI}}/t=3.83\pm0.05$ 
for the honeycomb lattice and 
$U_{\text{c}}^{\text{MI}}/t=5.56\pm0.06$ 
for the $\pi$-flux model~\cite{Note_eta}.
The critical values for $U_{\text{c}}^{\text{MI}}$ are thus consistent 
with those for $U_{\text{c}}$ determined from the AF order parameter $m_{\text{s}}$ 
within the statistical errors.
These results clearly imply that 
the transition from the SM to the AF insulator is continuous
in both models,
where the insulating behavior shows up immediately when 
the AF order is developed for $U \ge U_{\text{c}}$, 
i.e., $U_{\text{c}}^{\text{MI}} = U_{\text{c}}$. 
Therefore, our large-scale calculations exclude the intermediate phases 
previously reported in Refs.~\cite{Meng_Nature2010} and \cite{Chang_PRL2012}, 
and reveal a continuous transition between the SM and the AF insulator. 
As shown in Sec.~\ref{sec:fss}, the careful finite-size scaling analysis finds that 
the data collapse fits for both $m_{\text{s}}$ and $Z$ are convincing
with setting $U_{\text{c}}^{\text{MI}} = U_{\text{c}}$, 
also suggesting that the quantum critical points for these two quantities 
are located at the same $U$ value. 
 
Finally, we remark a semantic issue. 
The ``Mott transition'' is very widely used to describe a metal-insulator 
transition driven by the electron correlation, regardless of whether the 
symmetry may or may not be broken in the insulating phase. 
There are certainly more confusion and ambiguity on how to define properly 
a ``Mott insulator'', especially in our cases when the unit cell contains 
an even number of electrons and the AF order found in the 
insulating phase does not break the translation symmetry. 
However, this semantic 
issue is irrelevant to the main purpose of our study and does not change our 
conclusions.

\section{\label{sec:fss} finite-size scaling analysis} 

Having established the continuous character of the transition, 
let us now evaluate the critical exponents which characterize the quantum 
phase transition. For this purpose, here we employ the careful finite-size 
scaling analysis for staggered magnetization and quasiparticle weight. 

\subsection{\label{subsec:fss_m}staggered magnetization}

For the staggered magnetization, 
we make use of the standard finite-size scaling ansatz~\cite{Cardy_1988,Beach_condmat2005,Campostrini_PRB2014},
 \begin{equation}
  m_{\text{s}}(u, L) = L^{-\beta/\nu} 
   \left( 1 + c L^{-\omega} \right)
   f_{m}(u L^{1/\nu}),
   \label{eq:fss_m}
 \end{equation}
where $\nu$ is the critical exponent of the correlation length 
$\xi \simeq |U-U_{\text{c}}|^{-\nu}$,
$f_{m}(u L^{1/\nu})$ denotes a model dependent scaling function
and $u=(U-U_{\text{c}})/U_{\text{c}}$ is the reduced coupling.
Notice that the $u$ dependence of $m_{\text{s}}(L)$ given in 
Eq.~(\ref{eq:m_s}) is explicitly indicated here in Eq.~(\ref{eq:fss_m}) as $m_{\text{s}}(u, L)$. 
In the above finite-size scaling ansatz, 
we also take into account the leading correction term, 
a term proportional to $c L^{-\omega}$, with $c$ and $\omega$ 
being additional fitting parameters~\cite{note_omega},
which is however expected less important for sufficiently large cluster sizes. 
The finite-size scaling analysis is performed
with a recently proposed method based on the 
Bayesian statistics~\cite{Harada_PRE2011}.
The remarkable advantages of this method are 
i)  the weak dependence on the initial fitting parameters 
and 
ii) the applicability to a wide range of the reduced coupling $u$.

\begin{figure}[htbp]
 \centering
 \includegraphics[width=0.48\textwidth]{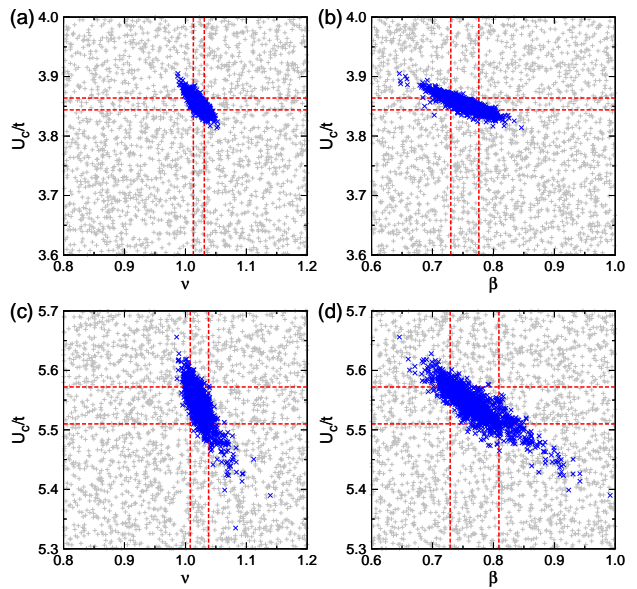}
 \caption{\label{fig:scatter}%
 Scattering plots of the computed fitting parameters (blue crosses) 
 for $m_{\text{s}}(u,L)$ obtained with the Bayesian method~\cite{Harada_PRE2011} 
 and the resampling procedure described in the text.  
 Here the data collapse fits for $m_{\text{s}}(u,L)$ 
 with the leading correction term given in Eq.~(\ref{eq:fss_m}) 
 are performed several thousands times: 
 we generate different input data sets (i.e., resampled data sets) that are 
 statistically consistent with the raw QMC data of $m_{\text{s}}(u,L)$, 
 and employ the Bayesian method 
 to perform the data collapse fit for each resampled data set with a different set 
 of initial fitting parameters (gray symbols) prepared randomly. 
 We use  $L \ge 15$ for the honeycomb lattice model [(a) and (b)] 
 and $L \ge 20$ for  the $\pi$-flux model [(c) and (d)].
 The vertical and horizontal dashed lines indicate 
 one standard deviation from the averaged 
 values of the computed fitting parameters. 
 }
\end{figure}

\begin{figure}[htbp]
 \centering
 \includegraphics[width=0.48\textwidth]{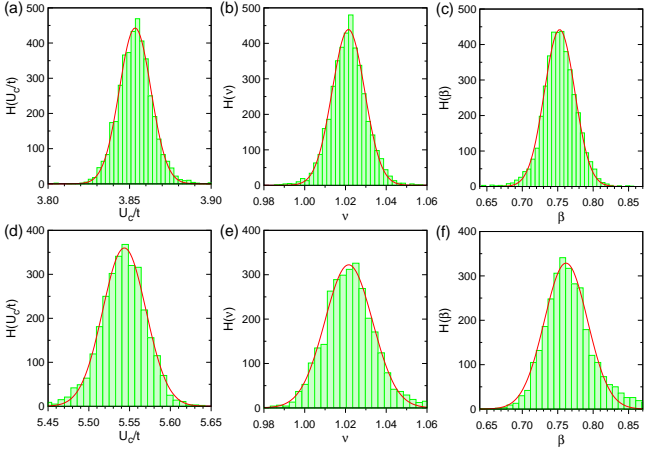}
 \caption{\label{fig:hist}%
 Histograms of the computed fitting parameters for $m_{\text{s}}(u,L)$ 
 obtained with the Bayesian method~\cite{Harada_PRE2011} and the resampling 
 procedure described in the text. 
 The same data sets as in Fig.~\ref{fig:scatter} are used 
 for (a)--(c) the honeycomb lattice model
 and for (d)--(f) the $\pi$-flux model.
 Solid curves are fits to the histograms with a Gaussian function. 
 These results clearly show that, after performing the Bayesian analysis several 
 thousands times for different resampled data sets generated 
 statistically consistent with the raw QMC data,  
 the computed fitting parameters are distributed normally 
 and therefore we can reasonably estimate 
 their average values and the corresponding statistical errors. 
 } 
\end{figure}

In order to estimate reliable error bars 
for the fitting parameters in Eq.~(\ref{eq:fss_m}), 
we adopt a straightforward resampling technique. 
The fitting procedure is summarized as follows. 
First, we prepare a data set to be fitted, 
based on the raw QMC data of $m_{\text{s}}(u,L)$ for 
various $u$ and $L$, by adding to $m_{\text{s}}(u,L)$ 
a Gaussian-distributed noise with the zero average and 
the standard deviation estimated by the QMC calculations for $m_{\text{s}}(u,L)$.
Second, we pick up at random 
initial values of the fitting parameters 
$U_{\text{c}}/t$, $\nu$, $\beta$, and $\omega$ 
around the optimal ones, namely, 
$U_{\text{c}}/t=3.8$ $(5.5)$ for the honeycomb lattice ($\pi$-flux) model,
$\nu=1.0$, $\beta=0.8$, and $\omega=0.8$. 
Third, with these initial fitting parameters, 
we perform the data collapse analysis for each resampled data for $m_{\text{s}}(u,L)$ 
based on the Bayesian statistics~\cite{Harada_PRE2011} 
and obtain the best converged values of 
$U_{\text{c}}/t$, $\nu$, $\beta$, and $\omega$. 
We repeat this procedure typically a few thousands times 
to average the converged parameters and estimate 
the statistical errors. 
Typical examples of the resampling procedure are shown in 
Figs.~\ref{fig:scatter} and~\ref{fig:hist}.
The clear advantage of the resampling technique is that 
the degree of uncertainty of the fitting parameters
can be immediately verified 
and therefore a reliable estimate of the error bars can be 
safely obtained. 

\begin{figure}[ht]
 \centering
 \includegraphics[width=0.48\textwidth]{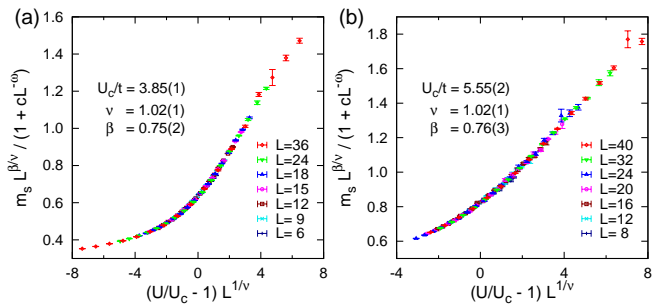}
 \caption{\label{fig:collapse_m}%
 Data collapse fits of the staggered magnetization $m_{\text{s}}$ 
 for (a) the honeycomb lattice model 
 and (b) the $\pi$-flux model. 
 The scaling form is given in Eq.~(\ref{eq:fss_m}), 
 where the fitting parameters 
 $U_{\text{c}}$, $\nu$, $\beta$, $c$, and $\omega$ are determined 
 by the resampling technique 
 using the data sets for $L\ge15$ in (a) and $L\ge20$ in (b) 
 (see also Table.~\ref{tbl:m_with} in Appendix~\ref{appsec:correction}).
 The $U_\text{c}$, $\mu$ and the system sizes $L$ used 
 are indicated in the figures. 
 The number in parentheses denotes the estimated error 
 in the last digit.
 }
\end{figure}

The results of the data collapse for the staggered magnetization $m_{\text{s}}(u,L)$ 
are shown in Fig.~\ref{fig:collapse_m}, 
confirming that our numerical calculations are quite accurate for this
quantity since the data for different $L$ collapses almost perfectly 
into a universal curve. 
We find that the critical exponents are quite stable
and converged to $\nu \simeq 1 $ and $\beta \simeq 0.75$ for both models, 
as indicated in Fig.~\ref{fig:collapse_m}.
It should be noticed that, for both models, 
the values of $U_{\text{c}}$, 
obtained from the data collapse plots in Fig.~\ref{fig:collapse_m}, 
agree within two standard deviations with the ones estimated 
straightforwardly by extrapolating $m_{\text{s}}(L)$ to 
the thermodynamic limit for each $U$ (see Figs.~\ref{fig:m2vsL} and \ref{fig:Uc}).

\begin{figure}[ht]
 \centering
 \includegraphics[width=0.48\textwidth]{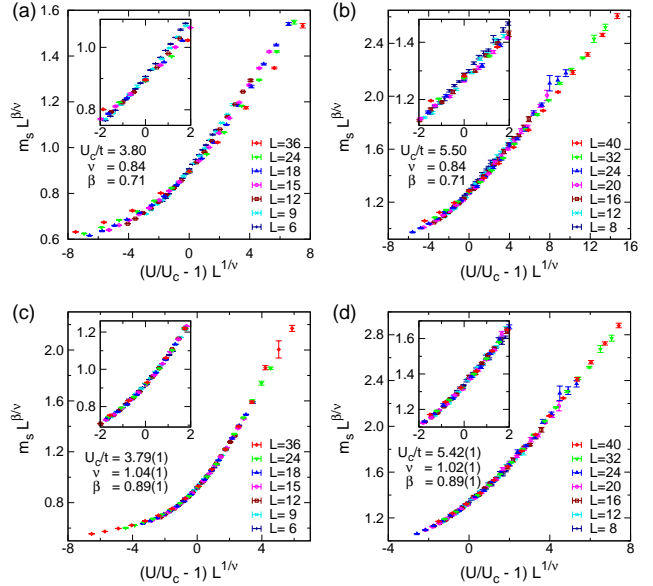}
 \caption{\label{fig:collapse_fa}%
 Data collapse plots of the staggered magnetization $m_{\text{s}}$ 
 for (a) the honeycomb lattice model 
 and (b) the $\pi$-flux model 
 with fixed values of $U_{\text{c}}$, $\nu$, and $\beta$, 
 taken from Ref.~\cite{ParisenToldin_PRB2015}. 
 In Ref.~\cite{ParisenToldin_PRB2015}, the critical exponents  $\nu$ and $\beta$ 
 are estimated, based on their numerical data and renormalization group analysis, 
 only for the honeycomb lattice model, and are assumed to 
 be the same for the $\pi$-flux model. 
 For comparison, data collapse fits without the leading correction term
 are also shown in the lower panels
 for (c) the honeycomb lattice model 
 and (d) the $\pi$-flux model, where 
 $U_{\text{c}}$, $\nu$, and $\beta$ are determined by the resampling technique 
 using the data sets for $L\ge9$ in (c) and $L\ge12$ in (d) 
 (see also Table.~\ref{tbl:m_without} in Appendix~\ref{appsec:correction}).  
 In all figures, the same data are used as those 
 in Fig.~\ref{fig:collapse_m}, but  
 the leading correction term is not considered, 
 i.e., $c=0$ in Eq.~(\ref{eq:fss_m}), for a fair comparison.
 }
\end{figure}

It should be noted here that very recently the similar QMC method has been 
applied to the honeycomb lattice model and the $\pi$-flux model 
by Parisen Toldin \textit{et al.} in Ref.~\cite{ParisenToldin_PRB2015}. 
In their report, the critical exponents for the honeycomb lattice model are 
estimated as $\nu \simeq 0.84$ and $\beta \simeq 0.71$, and are claimed to be 
consistent with those obtained by the $\epsilon$-expansion
for the GN model~\cite{Herbut_PRB2009a} (see also Table~\ref{tbl:exponents}), 
while the critical exponents for the $\pi$-flux model are unavailable.
We argue that 
the disagreement of the critical exponents between their estimations and ours
for the honeycomb model and the difficulty to determine 
the critical exponents for the $\pi$-flux model in Ref.~\cite{ParisenToldin_PRB2015} 
are due to the limited lattice sizes used in Ref.~\cite{ParisenToldin_PRB2015}, 
i.e., 
up to 648 sites for the honeycomb lattice model 
and 784 sites for the $\pi$-flux model. 
As shown in Figs.~\ref{fig:collapse_fa}(a) and \ref{fig:collapse_fa}(b), 
our numerical data do not provide perfect data collapse plots for both models 
when we fix $U_{\text{c}}$, $\nu$, and $\beta$ which are reported in 
Ref.~\cite{ParisenToldin_PRB2015}. 
However, the scattering of the data is particularly evident only 
for the two largest sizes, not studied in Ref.~\cite{ParisenToldin_PRB2015}.
Indeed, if these largest data sets are excluded from the plots, the data collapse seems 
acceptable even by using the values of $U_{\text{c}}$, $\nu$, and $\beta$ 
reported in Ref.~\cite{ParisenToldin_PRB2015},
especially for the honeycomb lattice model shown in Fig.~\ref{fig:collapse_fa}(a).  
On the other hand, Figs.~\ref{fig:collapse_fa}(c) and \ref{fig:collapse_fa}(d)  
confirm that our estimation of the critical values 
even without the leading correction term 
(see Appendix~\ref{appsec:correction} for the details)
yields an excellent data collapse.

\subsection{\label{subsec:fss_n}quasiparticle weight}

As far as the critical behavior in the charge sector is concerned,
the scaling ansatz is applied to the jump $\Delta n(u, L)$ of 
the momentum distribution function $n(\varepsilon_{\bm{k}})$
across the Fermi level with a form 
\begin{equation}
 \Delta n(u, L) = L^{-\eta_{\psi}} f_{n}(u L^{1/\nu}),
 \label{eq:fss_n}
\end{equation}
where $\eta_\psi$ is the anomalous dimension of 
the fermion field $\Psi$~\cite{Amit_1984}
and $f_{n}(u L^{1/\nu})$ is a scaling function. 
The jump $\Delta n(u,L)$ is simply obtained as a difference of 
$n(\varepsilon_{\bm{k}})$ at the two closest points to the Fermi level, 
i.e., the ones above and below the Fermi level (see Fig.\ref{fig:nk}).
Here we do not take $Z_L$ as a scaling quantity in Eq.~(\ref{eq:fss_n})
to avoid possible artifacts caused by the extrapolation procedure 
for $Z_{L}$ where $n(\varepsilon_{\bm{k}})$ is extrapolated to $\varepsilon_{\bm{k}}=0$.
The finite-size scaling for $\Delta n(u, L)$ is certainly more difficult 
because $\Delta n(u, L)$ is the direct finite size jump of $n(\varepsilon_{\bm{k}})$ across 
the Fermi level and is much larger than the quasiparticle weight $Z$ 
in the thermodynamic limit, as shown in Fig.\ref{fig:nk}.
Therefore, we fix the critical $U_{\text{c}}$ and
the exponent $\nu$ in Eq.~(\ref{eq:fss_n}) to the values already determined by the 
finite-size scaling analysis on $m_{\text{s}}(u,L)$.
Moreover, we do not consider the correction term in the finite size scaling analysis 
(see Appendix~\ref{appsec:correction} 
for the case with the leading correction term). 
The remaining parameter $\eta_{\psi}$ is determined using the resampling 
technique described in Sec.~\ref{subsec:fss_m}.

\begin{figure}[ht]
 \centering
 \includegraphics[width=0.48\textwidth]{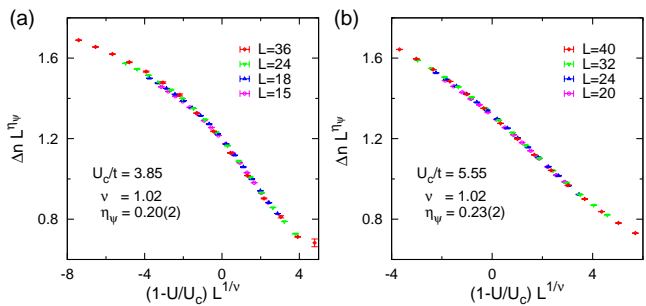}
 \caption{%
 Data collapse fits of $\Delta n(u, L)$ 
 for (a) the honeycomb lattice model 
 and (d) the $\pi$-flux model. 
 The scaling form is given in Eq.~(\ref{eq:fss_m}), 
 where $U_{\text{c}}$ and $\nu$ are fixed 
 to the values determined by the finite-size scaling analysis on the 
 staggered magnetization $m_{\text{s}}(u,L)$ (see Fig.~\ref{fig:collapse_m}). Thus,  
 only  $\eta_\psi$ is considered in the fitting procedure. 
 The critical exponents and the system sizes $L$ used 
 are indicated in the figures. 
 The number in parentheses indicates the estimated error 
 in the last digit. 
 }
 \label{fig:collapse_n}
\end{figure}

Despite the above simplifications, we find in Fig.~\ref{fig:collapse_n} that  
the collapse plots are excellent also in this case, 
suggesting that the lattice sizes considered here 
are large enough and that we can faithfully describe the critical 
behavior also in the charge sector. 
The obtained values for the critical exponent $\eta_\psi$ are found to be 
the same for both models ($\eta_\psi \simeq 0.21 - 0.22$) within statistical errors. 
However, these values differ from those obtained in Eq.~(\ref{eq:z_eta}) 
by taking the simple power law fit of $Z_L$ to extrapolate in the thermodynamic limit 
(see Figs.~\ref{fig:Uc} and \ref{fig:ZvsL})~\cite{Note_eta}.
We believe that the finite-size scaling analysis based on
Eq.~(\ref{eq:fss_n}) results in a more accurate estimation of the exponent,
since all data, not only in the metallic region for $U<U_{\text{c}}$ 
but also in the insulating region for $U>U_{\text{c}}$, are used.

The obtained critical exponents are summarized in Table~\ref{tbl:exponents} 
for the honeycomb lattice model and the $\pi$-flux model. 
As clearly shown in Table~\ref{tbl:exponents}, 
these two different models lead to the same critical exponents 
for the spin and charge sectors with a considerable degree of accuracy. 
Therefore, our numerical results firmly verify 
the universal quantum criticality 
in the apparently different lattice models, 
which share only the massless Dirac energy dispersion in the non-interacting limit. 
To our knowledge, 
this represents one of the first unbiased studies 
on the critical properties of 
the metal-insulator transition in two spatial dimensions. 
The highly accurate estimation of the critical exponents 
as well as the critical quantities follows from the unprecedentedly 
large-scale simulations that we are now able to perform~\cite{Sorella_SR2012}, 
combined with the careful and accurate finite size scaling.

\subsection{\label{subsec:superposed}scaling functions}

\begin{figure}[ht]
 \centering
 \includegraphics[width=0.48\textwidth]{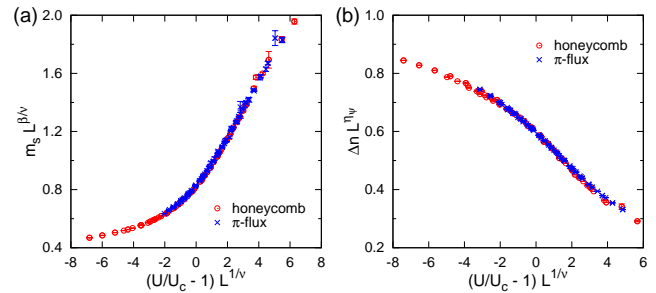}
 \caption{%
 Superposed scaling functions 
 of the honeycomb lattice model (red circles) 
 and the $\pi$-flux model (blue crosses)
 for (a) the staggered magnetization
 and (b) the jump in $n(\varepsilon_{\bm{k}})$.
 Data for the $\pi$-flux model are scaled with
 non-universal metric factors 
 $c_{1}=0.768$ and $c_{2}=1.38$ for (a)
 and 
 $c_{1}=0.851$ and $c_{2}=1.10$ for (b).
 The system sizes used are
 $L\ge 15$ for the honeycomb lattice model 
 and $L\ge 20$ for the $\pi$-flux model.
 }
 \label{fig:superposed}
\end{figure}

In order to firmly establish the universal character of 
the metal-insulator transition, 
we follow the Privman and Fisher's argument~\cite{Privman_PRB1984}. 
It states that if two different models belong to the same universality class, 
the scaling functions of a physical quantity for these two models are 
related by non-universal metric factors as
\begin{equation}
 f_{\alpha}(x) = c_{2} f_{\alpha}^{\prime}(c_{1} x),
 \label{eq:privman}
\end{equation}
where $f_{\alpha}(x)$ and $f^{\prime}_{\alpha}(x)$ are
the scaling functions for the two models.
Figures~\ref{fig:superposed}(a) and \ref{fig:superposed}(b) show 
the collapse fits without the correction term for the staggered magnetization
and for the jump in $n(\varepsilon_{\bm{k}})$, respectively.
We clearly find in Fig.~\ref{fig:superposed} that 
the collapse fits for the honeycomb lattice model can superpose onto 
the appropriately scaled collapse fits for the $\pi$-flux model with the non-universal 
constants $c_1$ and $c_2$. 
This implies that the scaling functions of the two models are essentially equal 
and confirms in a robust and unambiguous way the existence of the universal 
critical behavior of the metal-insulator transition studied here.

\section{\label{sec:vf} Charge structure factor}

Let us now investigate the long wavelength limit or 
equivalently the small $|\bm{q}|$ behavior of the 
static charge structure factor $N({\bm q})$. 
Being the static structure factor $N({\bm q})$ the integral over all the frequencies of 
the dynamical structure factor $N({\bm q},\omega)$,
i.e. $N({\bm q})=\int d\omega N({\bm q},\omega)$,
it depends, e.g., as in the spatial dimensionality $D>1$ 
within Fermi liquid theory~\cite{pines,Polini_SSC2007}, 
on the charge and the Fermi velocities, 
as well as on other low energy parameters such as the Landau parameter $F_0^s$ 
in the standard Fermi liquid theory.
This is therefore an interesting quantity and 
gives us information on how the dynamics evolves 
as we approach the metal-insulator transition point 
at $U_{\text{c}}$ from the metallic side. 
The static charge structure factor $N({\bm{q}})$ at momentum $\bm{q}$ is defined as 
\begin{equation}
N(\bm{q})=\frac{1}{{N_{\text{U}}}} 
\sum_{j,k=1}^{N_{\text{U}}} e^{i \bm{q}\cdot (\bm{r}_{j}-\bm{r}_{k})}
\langle  n_j n_k  \rangle, 
\end{equation}
where $N_{\text{U}}$ is the number of unit cells, 
$n_j=\sum_\alpha\sum_s c^\dag_{j\alpha s}c_{j\alpha s}$, and 
$c^\dag_{j\alpha s}$ is the creation operator of electron at unit cell ${\bm r}_j$ and 
sublattice $\alpha$ ($=A,B$) with spin $s\,(=\uparrow,\downarrow)$.
In the metallic region for $U < U_{\text{c}}$, 
the charge structure factor should behave 
as the non-interacting one, i.e., 
\begin{equation}
N(\bm{q}) \sim \alpha  |\bm{q}|^2 \ln |\bm{q}| 
\label{eq:Nq1}
\end{equation} 
for small $|\bm{q}|$, 
where $\alpha$ is a suitable constant 
that can be renormalized by the interaction, 
as in Fermi liquid theory~\cite{pines}.

\begin{figure}[ht]
 \centering
 \includegraphics[width=0.40\textwidth]{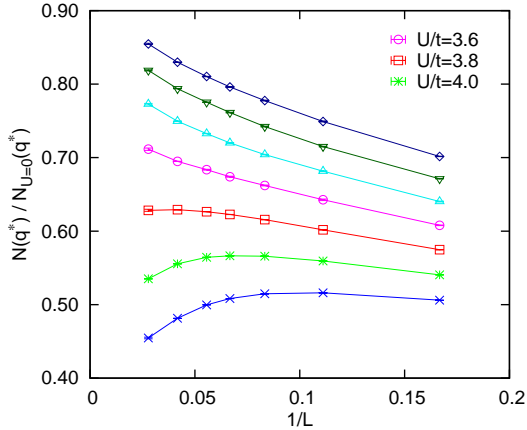}
 \caption{\label{fig:Nq-L}%
 Charge structure factors $N(\bm{q})$ 
 at the smallest non-zero momentum $|\bm{q}|=q^{\ast}$
 divided by the charge structure factor $N_{U=0}(q^{\ast})$ in the non-interacting limit 
 for the honeycomb lattice model.
 The interaction parameters are $U/t$ = 3.0, 3.2, 3.4, 3.6, 3.8, 4.0, and 4.2
 from top to bottom.
 }
\end{figure}

Figure~\ref{fig:Nq-L} shows the ratio 
\begin{equation}
R=\frac{N(q^{\ast})}{N_{U=0}(q^{\ast})} 
\end{equation} 
of the charge structure factor for finite $U$ and the noninteracting limit, 
denoted as $N_{U=0}({\bm q})$, 
at the smallest non-zero momentum $q^{\ast}$ available for a given system size. 
Since $N({\bm q})\simeq |{\bm q}|^2$ in the insulating 
phase~\cite{Feynman_PR1956,Capello_PRL2005,Capello_PRB2006}, 
we expect that $R\sim1/\ln L$ in the insulating phase, 
whereas $R \sim \text{const.}$ in the metallic phase. 
Therefore, there is a change of behavior in $R$ across  
the critical value of $U_{\text{c}}$. 
Indeed, we find in Fig.~\ref{fig:Nq-L} 
that R decreases with increasing $L$ for $U$ larger than $U_{\text{c}}$.

\begin{figure}[ht]
 \centering
 \includegraphics[width=0.40\textwidth]{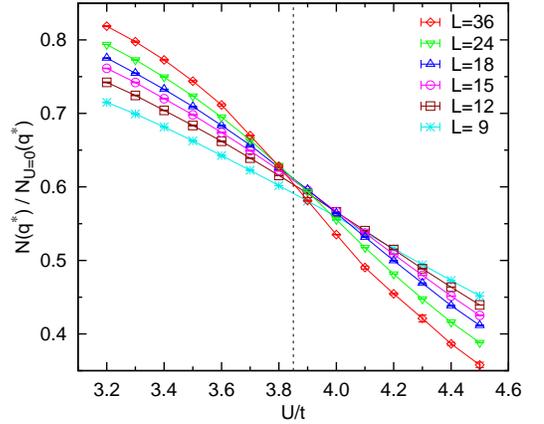}
 \caption{\label{fig:Nq}%
 Charge structure factors $N(\bm{q})$ 
 at the smallest non-zero momentum $|\bm{q}|=q^{\ast}$ 
 divided by the the charge structure factor $N_{U=0}(q^{\ast})$ in the non-interacting limit 
 for the honeycomb lattice model with different system sizes indicated in the figure. 
 The dashed line denotes the critical $U_{\text{c}}$ 
 determined by the data collapse 
 fit of the staggered magnetization in Fig.~\ref{fig:collapse_m}(a). 
 }
\end{figure}

It should be emphasized that $R$ remains finite as we approach 
the transition point from the metallic side. 
As shown in Fig.~\ref{fig:Nq}, $R$ as a function of $U/t$ 
for different system sizes crosses around the critical point 
at a {\it finite} value of $R$,  
which indicates that the coefficient $\alpha$ in Eq.~(\ref{eq:Nq1})
is neither singular nor critical at the critical point for $U \to  U_{\text{c}}$, 
but approaches a finite constant in the thermodynamic limit. 
On the other hand, this coefficient $\alpha$ 
vanishes at the metal-insulator transition 
within the Gutwiller~\cite{Gutzwiller_1965} 
or Brinkman-Rice~\cite{Brinkman_PRB1970} approximation 
on any lattice (including the honeycomb lattice) 
as a result of the vanishing of the double occupancy
$\bar{d}=\langle n_{i\uparrow} n_{i\downarrow} \rangle $
and the sum rule valid at half-filling 
\begin{equation} 
\sum_{|{\bm{q}}|\ne 0} N({\bm{q}}) = N_{\text{U}} (4 \bar{d} + 2  g_{\text{NN}}), 
\label{thebound}
\end{equation}
where $g_{\text{NN}}$ is the nearest neighbor density correlation, 
i.e., $g_{\text{NN}}= \langle n_{jA} n_{jB} \rangle - \langle n_{jA}\rangle \langle n_{jB} \rangle $, 
and $n_{j\alpha}=\sum_s c^\dag_{j\alpha s} c_{j\alpha s}$. 
This is simply because $g_{\text{NN}} \to 0$ when $\bar d\to 0$ 
at the metal-insulator transition described by the Gutzwiller approximation~\cite{note_nq}, 
and is therefore in sharp contrast with our results.

Although our results do not represent a direct evidence for the non-singular 
behavior of $v_{\text{F}}$ close to the metal-insulator transition, 
$v_{\text{F}}$ is certainly related to the charge velocity, as in Fermi liquid theory, 
which in turn affects the value of $\alpha$. 
Therefore, our results imply that $v_{\text{F}}$ remains finite 
at the metal-insulator transition. 
In this respect, it should be noted that 
the evolution of $\alpha$ as a function of $U$ found in Fig.~\ref{fig:Nq} 
is compatible with the expected behavior of the Dirac Fermi velocity $v_{\text{F}}$
for the GN universality class of the metal-insulator transition~\cite{Herbut_PRB2009b}.
As discussed previously~\cite{Wu_PRB2014},
the vanishing of the quasiparticle weight $Z$ 
without a renormalization of $v_{\text{F}}$ 
is understood as a consequence of 
an equal divergence in the momentum $\bm{k}$ and frequency $\omega$ 
derivatives of the electron self-energy $\Sigma({\bm{k}},\omega)$ 
at the Dirac point. 
The quasiparticle weight $Z$ at the Dirac point with momentum ${\bm{k}}_{\text{F}}$ 
is given as 
\begin{equation}
Z = \left[  1 - \left. 
\frac{\partial}{\partial\omega}{\text{Re}}\Sigma({\bm{k}}_{\text{F}},\omega) 
\right|_{\omega=0} \right]^{-1},
\end{equation}
while the Dirac Fermi velocity $v_{\text{F}}$ is related to $Z$ as 
\begin{equation}
\frac{v_{\text{F}}}{v_{\text{F}}^{0}} = 
Z \left[ 1 + \frac{1}{v_{\text{F}}^{0}} \left. 
\frac{\partial}{\partial k} {\text{Re}}\Sigma({\bm{k}},0)
\right|_{{\bm{k}}={\bm{k}}_{\text{F}}} \right]. 
\end{equation}
Therefore, in order to compensate the divergence of 
$\left. \frac{\partial}{\partial\omega}{\text{Re}}\Sigma({\bm{k}}_{\text{F}},\omega) \right|_{\omega=0} $, 
i.e., $Z\to0$, at the metal-insulator transition point, 
$ \left. \frac{\partial}{\partial k} {\text{Re}}\Sigma({\bm{k}},0)\right|_{{\bm{k}}={\bm{k}}_{\text{F}}}$ must diverge at the same time. 
This implies that the strong momentum dependence of $\Sigma({\bm{k}},\omega)$ 
around the Dirac point, 
not included in the simplest DMFT approach, is an essential ingredient 
to describe the metal-insulator transition.  
Similar arguments are also found in earlier studies of the $t$-$J$ model
with the large-$N$ expansion~\cite{Kotliar_lecture1991}.
It is also interesting to note that the nodal Fermi velocity remains finite 
in the carrier number controlled Mott metal-insulator transition described 
by a Gutzwiller projected $d$-wave BCS state, 
which exhibits the massless Dirac dispersion at the nodal point~\cite{Yunoki_2005}.

In any event, our results are certainly useful to characterize the metal-insulator transition 
in the charge sector. 
Indeed, the critical value $U_{\text{c}}$ in the charge sector 
can be estimated directly by this simple analysis shown in Fig.~\ref{fig:Nq} 
without performing 
rather elaborated finite size scaling of quantities such as 
the charge gap~\cite{Meng_Nature2010,Chang_PRL2012}, 
which is usually very time consuming and difficult 
to compute with high accuracy.

\section{\label{sec:conclusion}Discussion and Conclusions}

The GN models have been extensively studied in quantum field theory~\cite{Gross_PRD1974}.
In the standard field theoretical treatment of the transition, 
the critical behavior is studied in space-time dimension $d=D+1$, 
where $D$ is the spatial dimensionality. 
The critical exponents of the GN models have been evaluated by several standard techniques, 
such as the large $N$ expansion~\cite{Gracey_1994,Karkkainen_NuclPhysB1994} and 
the $\epsilon$-expansion around the lower $d=2+\epsilon$~\cite{Gracey_NuclPhysB1990} 
or upper $d=4-\epsilon$~\cite{Rosenstein_PhysLettB1993} critical dimensions, 
and these are summarized in Table~\ref{tbl:exponents}. 
It is clearly noticed in Table~\ref{tbl:exponents} that 
there exist sizable discrepancies among the critical exponents calculated 
by these different analytical techniques. 
Therefore, an unbiased numerical study is highly desired to clarify 
the critical behavior of the GN models.

\begin{table*}[ht]
 \caption{\label{tbl:exponents}%
 Critical exponents, $\nu$, $\beta$, and $\eta_{\psi}$, of 
 the interaction-driven phase transition in interacting Dirac fermions in $d=2+1$ 
 for the lattice models (honeycomb lattice and $\pi$-flux models)
 and the effective continuous models (Gross-Neveu models) 
 with the total number $N$ of fermion components. 
 Different classes correspond to different symmetries broken in the ordered phases. 
 Numbers in parentheses for $\nu$, $\beta$, and $\eta_{\psi}$ indicate statistical errors in the last digits. 
 For comparison, the critical exponents for other related models with $N=4$ 
 and $8$, belonging to different universality classes, are also listed. 
 FRG stands for the functional renormalization group method. 
 }
 \begin{ruledtabular}
 \begin{tabular}{l l l l l l l}
  model              &
  $N$     &
  class              &   %
  method             &   %
  $\nu$              & 
  $\beta$            & 
  $\eta_{\psi}$      \\  
  \hline
  honeycomb          &
  8                  &
  chiral Heisenberg  &
  Monte Carlo (present)       &
  1.02(1)            &
  0.76(2)            &
  0.20(2)            \\
  $\pi$-flux         &
  8                  &
  chiral Heisenberg  &
  Monte Carlo (present) &
  1.02(1)            &
  0.74(3)            &
  0.23(2)            \\
  honeycomb          &
  8                  &
  chiral Heisenberg  &
  Monte Carlo \cite{ParisenToldin_PRB2015} &
  0.84(4)            &
  0.71(8)            &
  ---                \\
  %
  Gross-Neveu        &
  8                  &
  chiral Heisenberg  &
  $4-\epsilon$, 1st order \cite{Rosenstein_PhysLettB1993,note_Nf}  &
  0.851              &   
  0.804              &   
  0.167              \\  
  Gross-Neveu        &
  8                  &
  chiral Heisenberg  &
  $4-\epsilon$, 2nd order \cite{Rosenstein_PhysLettB1993,note_Nf}  &
  1.01               &   
  0.995              &   
  0.101              \\  
  Gross-Neveu        &
  8                  &
  chiral Heisenberg  &
  FRG \cite{Janssen_PRB2014} &
  1.31               & 
  1.32               &   
  0.08               \\  
  \\
  Gross-Neveu        &
  8                  &
  chiral Heisenberg  &
 $4-\epsilon$, 1st order \cite{Herbut_PRB2009a,note_Nf}&
  0.882              &   
  0.794              &   
  0.3                \\   
  \\
  Gross-Neveu        &
  4                  &
  chiral Heisenberg  &
  $4-\epsilon$, 1st order \cite{Rosenstein_PhysLettB1993}&
  0.882              &   
  0.794              &   
  0.3                \\  
  Gross-Neveu        &
  4                  &
  chiral Heisenberg  &
  $4-\epsilon$, 2nd order \cite{Rosenstein_PhysLettB1993}&
  1.083              &   
  1.035              &   
  0.242              \\  
  \\
  Gross-Neveu        &
  8                  &
  chiral Ising       &
  Monte Carlo  \cite{Karkkainen_NuclPhysB1994}&
  1.00(4)            &  
  0.88(4)            &   
  ---                \\ 
  Gross-Neveu        &
  8                  &
  chiral Ising       &   
  Monte Carlo  \cite{Chandrasekharan_PRD2013}&
  0.83(1)            &  
  0.67(1)            &   
  0.38(1)            \\ 
  Gross-Neveu        &
  8                  &
  chiral Ising       &
  $4-\epsilon$, 1st order \cite{Rosenstein_PhysLettB1993}  &
  0.738              &   
  0.631              &   
  0.071              \\  
  %
  %
  %
  Gross-Neveu        &
  8                  &
  chiral Ising       &
  $4-\epsilon$, 2nd order \cite{Rosenstein_PhysLettB1993}  &
  0.850              &   
  0.722              &   
  0.065              \\  
  Gross-Neveu        &
  8                  &
  chiral Ising       &
  $2+\epsilon$, 3rd order \cite{Gracey_NuclPhysB1990}  &
  1.309              &   
  1.048              &
  0.081              \\
  Gross-Neveu        &
  8                  &
  chiral Ising       &
  $O(1/N^2)$  \cite{Karkkainen_NuclPhysB1994,Gracey_1994,Vasilev_1993}&
  0.829              &   
  0.723              &   
  0.044              \\  
  Gross-Neveu        &
  8                  &
  chiral Ising       &
  FRG \cite{Rosa_PRL2001,Hofling_PRB2002} &
  1.018              &  
  0.894              &  
  0.032              \\ 
  Gross-Neveu        &
  8                  &
  chiral Ising       &
  FRG \cite{Janssen_PRB2014} &
  1.018              &  
  0.896              &  
  0.032              \\ 
  \\
  honeycomb          &
  4                  & 
  chiral Ising       &
  Monte Carlo \cite{Wang_NJP2014} &
  0.80(3)            &
  0.52(2)            &
      ---               \\
  $\pi$-flux         &
  4                  &
  chiral Ising       &
  Monte Carlo \cite{Wang_NJP2014} &
  0.80(3)            &
  0.53(4)            &
   ---                \\
  honeycomb          &
  4                  & 
  chiral Ising       &
  Monte Carlo \cite{Li_NJP2015} &
  0.60(3)            &
  0.77(3)            &
  ---                \\
  $\pi$-flux         &
  4                  &
  chiral Ising       &
  Monte Carlo \cite{Li_NJP2015} &
  0.67(4)            &
  0.79(4)            &
  ---                \\
  Gross-Neveu &
  4                  &
  chiral Ising       &
  $4-\epsilon$, 1st order \cite{Rosenstein_PhysLettB1993}  &
  0.709       &   
  0.559       &   
  0.100       \\  
  Gross-Neveu        &
  4                  &
  chiral Ising       &
  $4-\epsilon$, 2nd order \cite{Rosenstein_PhysLettB1993}  &
  0.797              &   
  0.610              &   
  0.110              \\  
  Gross-Neveu        &
  4                  &
  chiral Ising       &
  FRG \cite{Rosa_PRL2001,Hofling_PRB2002} &
  0.927              &  
  0.707              &  
  0.071              \\ 
  \\
  Gross-Neveu        &
  8                  &
  chiral XY          &
  $4-\epsilon$, 1st order \cite{Rosenstein_PhysLettB1993}  &
  0.726              &   
  0.619              &   
  0.071              \\  
  Gross-Neveu        &
  8                  &
  chiral XY          &
  $4-\epsilon$, 2nd order \cite{Rosenstein_PhysLettB1993}  &
  0.837              &   
  0.705              &   
  0.063              \\  
  \\
  Gross-Neveu        &
  4                  &
  chiral XY          &
  $4-\epsilon$, 1st order \cite{Rosenstein_PhysLettB1993}  &
  0.7                &   
  0.55               &   
  0.1                \\  
  Gross-Neveu        &
  4                  &
  chiral XY          &
  $4-\epsilon$,  2nd order \cite{Rosenstein_PhysLettB1993}  &
  0.799              &   
  0.607              &   
  0.106              \\  
\end{tabular}
\end{ruledtabular}
\end{table*}

In this paper, we have established, 
based on robust and reliable numerical simulations on fairly large clusters, 
the universal properties of the metal-insulator transition 
for the two different lattice models of interacting Dirac electrons 
in two spatial dimensions. 
We have determined the critical exponents which characterize
the universal quantum critical behavior
in both metallic and insulating phases at the vicinity of 
the transition for these models. 
Since it is expected that the effective low-energy theory of these lattice models are
described in the continuous limit by the chiral Heisenberg universality class 
of the GN model with $N=8$, 
our study represents currently the most accurate and reliable results
also for this fundamental GN model to reveal the universal critical behavior.
Indeed, our results resolve some of the inconsistency among different 
approximate approaches for the GN models
shown in Table~\ref{tbl:exponents}, especially evident for $N=8$. 

We have also clarified how the quasiparticle in the SM phase collapses 
as the AF insulating phase is approached with increasing $U$ for these two models. 
We have shown that the quasiparticle weight $Z$ diminishes and becomes zero at the 
metal-insulator transition and found that 
the exponent $\eta_{Z} =\nu \eta_{\psi} $, characterizing the renormalization 
of the quasiparticle weight $Z$ [see Eq.~(\ref{eq:z_eta})], 
is much smaller for both models ($\eta_{Z}  \simeq 0.2$) 
than the one predicted by simple mean-field and dynamical mean-field correlated theories 
of the metal-insulator transition, for which $\eta_{Z} =1$~\cite{Brinkman_PRB1970,Rosenberg}.
More interestingly, we have also found that 
small $q$ limit of the static charge structure factor is not singular 
at the metal-insulator transition, suggesting that the Dirac Fermi velocity 
$v_{\text{F}}$ is not critical at the transition. 
These critical behaviors, small $\eta_{Z}$ and finite $v_{\text{F}}$,
are qualitatively the same as the ones expected for the GN universality 
class of the metal-insulator transition~\cite{Herbut_PRB2009b}.
Therefore, our results provide a clear numerical evidence that 
the critical behavior of the GN model applies also to lattice models and 
describes correctly the metal-insulator transition in interacting Dirac electrons. 
It should be noted here that the non-criticality of $v_{\text{F}}$ for the honeycomb lattice model  
is also found by quantum cluster methods~\cite{Wu_PRB2010,Wu_PRB2014}, although 
it was first overlooked within the single-site DMFT~\cite{Tran_PRB2009}.

It should also be remarked that,
strictly speaking, the metal-insulator transition studied here
does not describe a genuine  ``Mott transition'' 
because the insulating phase breaks $SU(2)$ symmetry for $U > U_{\text{c}}$. 
Indeed, as discussed below, 
each type of possible symmetry breaking in interacting Dirac fermions 
determines a different universality class of the transition characterized 
with different critical exponents. 
Nevertheless, the metal-insulator transition studied here does \textit{not} 
originate from a Slater-type nesting instability at weak coupling 
since the density of states is zero at the Fermi level. 
The transition instead occurs at an intermediate or strong coupling region 
as evidenced by the fact that $U_{\text{c}}\simeq 5.55t$ for the $\pi$-flux model 
is almost the same as the non-interacting band width $4\sqrt{2}t$.
 
Very recently, using the continuous time QMC, 
Wang {\it et al.}~\cite{Wang_NJP2014} have studied 
similar models of spinless fermions 
with the nearest neighbor repulsion $V$, 
where, with increasing $V$, a transition from the SM to a staggered 
charge-density-wave (CDW) state occurs. 
With a finite size scaling analysis 
based on the results for clusters up to $450$ sites,
they have estimated the critical exponents, 
$\nu$ and $\beta$, for the CDW order parameter~\cite{Wang_NJP2014}. 
These critical exponents have been recently revisited by
the Majorana QMC method on larger clusters up to 1152 sites~\cite{Li_NJP2015}.
As seen in Table~\ref{tbl:exponents}, these exponents 
are clearly different from those for the spinful models that we have studied here. 
This is simply understood because 
these spinless and spinful models belong to different universality classes. 
Indeed, it is known that the spinless lattice models with the nearest neighbor 
interaction at half-filling are described in the continuous limit by the GN model 
with $N=4$ and the chiral Ising universality class. 
It is also interesting to notice in Table~\ref{tbl:exponents} that 
the critical exponents 
for the spinless lattice models estimated numerically are rather different 
from the analytical results. 
This also demonstrates that numerically exact studies are 
highly valuable to accurately determine the quantum criticality 
and to remove the ambiguities that might arise from inadequate approximations. 

It should be emphasized that the various universality classes depend on the physics, 
namely, the symmetry that is broken in the ordered phase 
and the total number $N$ of Dirac fermion components that describe 
the corresponding critical theory~\cite{Chandrasekharan_PRD2013}. 
The well explored universality classes 
for the GN models in the continuous limit include 
the following three classes: 
\begin{enumerate}
\item Chiral Ising universality class. 
$Z_2$ symmetry, i.e., a discrete order parameter is broken, 
for instance, 
when a commensurate CDW order settles. 
\item Chiral XY universality class. 
$U(1)$ symmetry is broken and 
the order parameter is characterized by an angle, 
as in a superconducting state. 
\item  Chiral Heisenberg universality class. 
$SU(2)$ symmetry is broken. 
This should occur in the transition studied here, 
as the order parameter---the staggered magnetization---is 
characterized by a vector with three components 
[$SU(2)$ is equivalent to $SO(3)$]. 
\end{enumerate}
Among these, the two classes have been studied so far 
based on the lattice models with unbiased numerical techniques:
the chiral Ising universality class 
by Wang {\it et al.}~\cite{Wang_NJP2014}
and Li {\it et al.}~\cite{Li_NJP2015},
and the chiral Heisenberg universality class 
studied here and in Ref.~\cite{ParisenToldin_PRB2015}.
We expect that the quantum criticality belonging to the 
chiral XY universality class emerges in a negative $U$ Hubbard model 
with the Dirac points of the non-interacting energy dispersion at the Fermi level, 
provided that the $SU(2)$ symmetry 
which relates the CDW order to the superconducting one
is not satisfied (otherwise the chiral Heisenberg universality class applies again). 
For example, by adding the next nearest-neighbor hopping $t^\prime$ 
in the same lattice models studied here but with a negative $U$ 
(no sign problem with $t^\prime$ in the negative $U$ Hubbard model),
the chiral XY universality class with different critical exponents
can be investigated in the same unbiased numerical approach.

In conclusions,
we have investigated the critical behaviors of the metal-insulator transition 
in the interacting Dirac electrons in two spatial dimensions, 
described by the two different lattice models, 
the Hubbard model on the honeycomb lattice and 
the $\pi$-flux Hubbard model on the square lattice, 
at half-filling. 
We have performed the unprecedentedly large-scale quantum Monte Carlo simulations
to systematically calculate the staggered magnetization and the momentum distribution function.
The calculation of the momentum distribution function is particularly important
because it allows us to examine the quasiparticle weight and thus explore
the critical behavior also in the metallic side,
which had never been successful previously in the unbiased numerically exact calculations.
The ground state phase diagrams determined by extrapolating the staggered magnetization
and the quasiparticle weight to the thermodynamics limit have revealed the continuous
nature of the transition between the SM and the AF insulator with no intermediate phase.
Therefore, our results firmly rule out the possibility of a spin liquid phase
for these two models proposed in the earlier studies.
We have obtained the critical exponents by the careful and accurate
finite-size scaling analysis and found that
the two lattice models belong to the same universality class.
Since the low-energy effective model in the continuous limit for these two models
is described by the GN model with $N=8$ and the chiral Heisenberg universality class,
our results represent currently the most accurate determination of the quantum
criticality of this universality class.
Finally, we have shown that the quasiparticle weight monotonically decreases with increasing $U$
and becomes zero at the metal-insulator transition, while the Fermi velocity seems non-critical.
This qualitatively important feature is indeed in good agreement with the one expected for the GN
universality class of the metal-insulator transition and
cannot be captured by the simple mean-field or Gutzwiller-type approximate argument.

\acknowledgements
We acknowledge B.~Rosenstein for useful discussions 
and for clarifying us the notations in Ref.~\cite{Rosenstein_PhysLettB1993}. 
We are also grateful to I.~F.~Herbut, A.~Gambassi, M.~Capone, K.~Harada, 
T.~Sato, and Y.~Hatsugai for valuable comments. 
This work has been supported in part by 
Grant-in-Aid for Scientific Research from MEXT Japan 
(under Grant Nos. 24300020 and 26400413),
RIKEN iTHES Project, MIUR-PRIN-2010, and 
MEXT HPCI Strategic Programs for Innovative Research (SPIRE) 
(Grant No. hp140168).
The numerical simulations have been performed on K computer 
at RIKEN Advanced Institute for Computational Science (AICS), 
RIKEN Cluster of Clusters (RICC), 
RIKEN supercomputer system (HOKUSAI GreatWave), and 
NEC SX-ACE at Cybermedia Center, Osaka University.

\appendix

\section{\label{appsec:nk_def}Energy resolved momentum distribution function}

The quasiparticle weight can be, in general, calculated from the jump in
the momentum distribution function at the Fermi level. 
For the honeycomb lattice model and the $\pi$-flux model studied here, 
the momentum distribution function $n_{\alpha \beta,s}(\bm{k})$ 
at momentum $\bm{k}$ is defined as 
\begin{equation}
 n_{\alpha \beta, s}(\bm{k}) = 
  \langle c^{\dag}_{\bm{k} \alpha s} c_{\bm{k} \beta s} \rangle, 
\end{equation}
where 
\begin{equation}
 c^{\dag}_{\bm{k} \alpha s} = \frac{1}{\sqrt{N_{\text{U}}}} \sum_{\bm{r}} 
 e^{-i \bm{k} \cdot \bm{r}} c^{\dag}_{\bm{r} \alpha s}, 
\end{equation}
$c^{\dag}_{\bm{r} \alpha s}$ is the creation operator of electron 
at unit cell $\bm{r}$ and sublattice $\alpha\, (=A,B)$ 
with spin $s\,=(\uparrow,\downarrow)$, 
and $N_{\text{U}}$ is the number of unit cells. 
Notice that, due to the particle-hole symmetry at half-filling, 
$n_{AA, s}(\bm{k}) = n_{BB,s }(\bm{k}) = 1/2$ for all momenta, 
and therefore only $n_{AB, s}(\bm{k})$ and $n_{BA, s}(\bm{k})$ 
are non trivial with $\left[n_{AB, s}(\bm{k})\right]^{\ast}=n_{BA, s}(\bm{k})$.

The ``energy resolved'' momentum distribution function 
$n(\varepsilon_{\bm{k}})$ is defined as 
\begin{equation}
n(\varepsilon_{\bm k}) = \langle \psi^{\dag}_{\bm{k},\pm,s} \psi_{\bm{k},\pm,s} \rangle,
\label{eq:nk}
\end{equation}
where the annihilation operators $\psi_{\bm{k}, -, s} $ and $\psi_{\bm{k}, +, s} $ of 
the bonding and anti-bonding states, respectively, are given as 
\begin{equation}
\psi_{\bm{k}, \pm, s} = \frac{1} {\sqrt{2}} 
\left( c_{\bm{k}As} \pm \frac{h_{\bm{k}}}{|h_{\bm{k}}|} c_{\bm{k}Bs} \right).
 \label{eq:non-interacting_orbitals}
\end{equation}
These states $\psi_{\bm{k}, \pm, s}$ are the eigenstates of the non-interacting Hamiltonian 
\begin{equation}
 H_0=\sum_{{\bm k},s}
 \begin{pmatrix}
  c^\dag_{{\bm k}As} &c^\dag_{{\bm k}Bs}
 \end{pmatrix}
 \begin{pmatrix}
  0 & h_{\bm k} \\
  h_{\bm k}^{\ast} & 0
 \end{pmatrix}
 \begin{pmatrix}
  c_{{\bm k}As}\\
  c_{{\bm k}Bs}
 \end{pmatrix}
\end{equation} 
at momentum $\bm{k}$ with the energy $\varepsilon_{\bm{k}}=\pm |h_{\bm{k}}|$. Here 
\begin{equation}
h_{\bm{k}} = -t( 1 + e^{-i \bm{k}\cdot\vec{\tau}_{1}} + e^{-i \bm{k}\cdot\vec{\tau}_{2}} )
\end{equation}
for the honeycomb lattice model and 
\begin{equation}
h_{\bm{k}} = -t (1 + e^{-i \bm{k}\cdot\vec{\tau}_{1}} + e^{-i \bm{k}\cdot\vec{\tau}_{2}} 
- e^{-i \bm{k}\cdot \left(\vec{\tau}_{1} + \vec{\tau}_{2}\right)}) 
\end{equation}
for the $\pi$-flux model, where  
$\vec{\tau}_{1}$ and $\vec{\tau}_{2}$ are the primitive translational vectors 
defined in Figs.~\ref{fig:lattice}(a) and \ref{fig:lattice}(b). 

\begin{figure}[ht]
 \centering
 \includegraphics[width=0.48\textwidth]{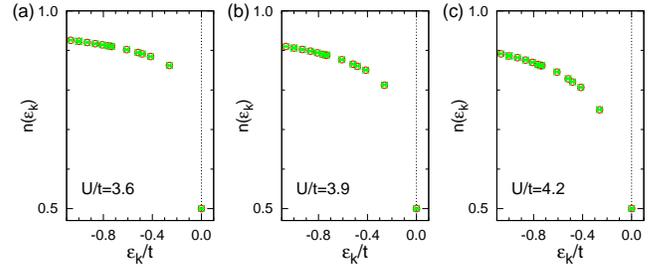}
 \caption{\label{fig:orbitals}%
 Energy resolved momentum distribution function 
 $n(\varepsilon_{\bm{k}})$ below the Fermi level
 ($\varepsilon_{\bm{k}}=0$) 
 for the honeycomb lattice model 
 with $L=24$ and different values of $U/t$ indicated in the figures. 
 Red circles and green crosses 
 represent  the results calculated using 
 the non-interacting bonding states [Eq.~(\ref{eq:non-interacting_orbitals})]
 and the natural orbitals [Eq.~(\ref{eq:natural_orbitals})], respectively.
 }
\end{figure}

In principle, in order to determine the quasiparticle weight $Z$,
the occupation number should be calculated in terms of the natural orbitals, 
i.e. the eigenvectors of the density matrix.
To this end, we should note that the density matrix at momentum $\bm{k}$ 
is a $2\times 2$ matrix and is represented as 
\begin{equation}
 \begin{pmatrix}
  n_{AA,s}({\bm k}) & n_{AB,s}({\bm k}) \\
  n_{BA,s}({\bm k}) & n_{BB,s}({\bm k})
 \end{pmatrix}
 =
 \begin{pmatrix}
  1/2               & f_{\bm{k}} \\
  f_{\bm{k}}^{\ast} & 1/2 
 \end{pmatrix}.
\end{equation} 
Therefore, the ``dressed'' quasiparticle operators simply read
\begin{equation}
 \bar\psi_{{\bm k},\pm,s}= \frac{1} {\sqrt{2}} 
  \left( c_{\bm{k}As} \pm \frac{f_{\bm{k}}^{\ast}}{|f_{\bm{k}}|} c_{\bm{k}Bs} \right) 
 \label{eq:natural_orbitals}
\end{equation}
with the occupation 
\begin{equation}
 \langle \bar\psi_{{\bm k},\pm,s}^\dag \bar\psi_{{\bm k},\pm,s}  \rangle 
  = \frac{1}{2} \pm  |f_{\bm{k}}|.
\end{equation} 
Notice that $f_{\bm k}=-\frac{1}{2}\frac{h_{\bm k}^{\ast}}{|h_{\bm k}|}$ 
in the non-interacting limit 
where the bonding (anti-bonding) states are all occupied (empty).
We have verified that even in the interacting case studied here 
the natural orbitals $\bar\psi_{\bm{k},\pm,s}$ are almost indistinguishable 
from the non-interacting bonding and anti-bonding states $\psi_{\bm{k},\pm,s}$
Indeed, we have found in Fig.~\ref{fig:orbitals} that the difference between 
the quasiparticle weight $Z$
calculated with the natural orbitals 
and the one with the non-interacting bonding and anti-bonding states 
is negligible for the models studied here both in the metallic and insulating phases. 
Therefore, the treatment for computing the quasiparticle weight $Z$ 
in Sec.~\ref{subsec:poorman_z} and Sec.~\ref{subsec:fss_n},
i.e, the jump of the energy resolved momentum distribution function 
$n(\varepsilon_{\bm{k}})$ given in Eq.~(\ref{eq:nk}) at the Fermi level, is 
not only asymptotically valid to determine the corresponding critical exponent 
but also represents a good quantitative estimate of $Z$.

\section{\label{appsec:correction}Effects of the leading correction term 
in the finite-size scaling analyses}

Here, we examine the robustness of the fitting parameters 
in Eqs.~(\ref{eq:fss_m}) and (\ref{eq:fss_n}) 
with and without the leading correction term in the finite-size scaling analyses. 
Since the leading correction term is expected less important for sufficiently 
large clusters, we examine the system size dependence of the fitting parameters. 

\begin{figure}[ht]
 \centering
 \includegraphics[width=0.48\textwidth]{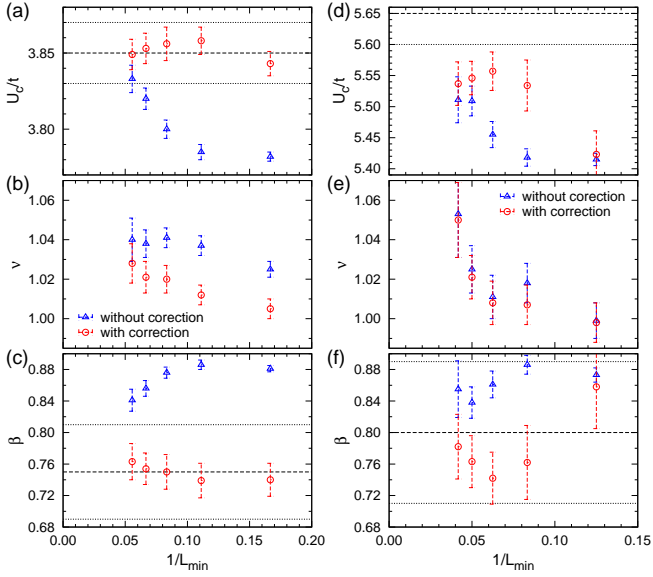}
 \caption{\label{fig:m-Lmin}%
 Same data as in Tables~\ref{tbl:m_without} and \ref{tbl:m_with}
 plotted as a function of $1/L_{\text{min}}$
 for the honeycomb lattice model (left panels)
 and the $\pi$-flux model (right panels).
 Red circles (blue triangles) represent the results
 with (without) the leading correction term in Eq.~(\ref{eq:fss_m}).
 For comparison,  $U_{\text{c}}/t$ and $\beta$ estimated directly 
 from the AF order parameter $m_{\text{s}}$ in Fig.~\ref{fig:Uc} 
 are indicated by dashed lines 
 (with the statistical errors denoted by dotted lines) 
 in (a), (c), (d), and (f). 
 Notice that these two different estimations 
 of $U_{\text{c}}/t$ and $\beta$ provide the statistically consistent results.
 }
\end{figure}

Tables~\ref{tbl:m_without} and \ref{tbl:m_with} summarize the fitting parameters in the 
staggered magnetization, which are obtained for clusters including the smallest lattice size 
$L_{\text{min}}$ in the data 
collapse without and with the leading correction term in the finite-size scaling ansatz 
of Eq.~(\ref{eq:fss_m}), respectively. 
These results are also compared in Fig.~\ref{fig:m-Lmin}
as a function of $L_{\text{min}}^{-1}$. 
It is noticed 
that $U_{\text{c}}/t$ systematically increases 
as the data sets with smaller $L$ are removed 
when we use the finite-size scaling ansatz without the leading correction term, 
i.e., $c=0$ in Eq.~(\ref{eq:fss_m}).
Accordingly, the critical exponent $\beta$ tends to decrease, 
although the critical exponent $\nu$ is less affected by 
including or not including the leading correction term 
in the finite-size scaling ansatz, 
at least, within the statistical errors.
These systematic differences with varying $L_{\text{min}}$ imply that 
the leading correction term in the finite-size scaling ansatz 
is not negligible. 

On the other hand, when the leading correction term is included,
the results are robust against the choice of $L_{\text{min}}$ 
as shown in Table~\ref{tbl:m_with} and Fig.~\ref{fig:m-Lmin}.
In addition, $U_{\text{c}}$ and $\beta$
evaluated in the data collapse analysis are 
statistically consistent with those obtained 
by the direct fit of thermodynamically extrapolated 
staggered magnetization, shown in Fig.~\ref{fig:Uc}. 
In Sec.~\ref{subsec:fss_m}, we report the results of the data collapses
with the leading correction term, 
because they are clearly more accurate and stable.
The quality of our data and extrapolations is further supported 
by the fact that 
both results obtained with and without the leading correction term 
tend to be identical when the system sizes are large enough, 
as shown in Tables~\ref{tbl:m_without} and \ref{tbl:m_with}, 
and in Fig.~\ref{fig:m-Lmin}.

\begin{figure}[ht]
 \centering
 \includegraphics[width=0.48\textwidth]{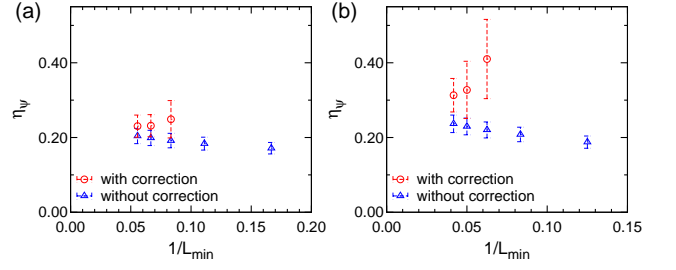}
 \caption{\label{fig:n-Lmin}%
 Same data as in Tables~\ref{tbl:n_without} and \ref{tbl:n_with}
 plotted as a function of $1/L_{\text{min}}$
 for (a) the honeycomb lattice model and (b) the $\pi$-flux model.
 Red circles (blue triangles) represent the results
 with (without) the leading correction term in Eq.~(\ref{eq:fss_n_lead}).
 }
 \end{figure}

The finite-size scaling ansatz for the jump of $n(\varepsilon_{\bm k})$ including 
the leading correction term is given as 
\begin{equation}
 \Delta n(u, L) = L^{-\eta_{\psi}} 
  \left( 1 + d L^{-\omega^{\prime}} \right)
  f_{n}(u L^{1/\nu} ),
 \label{eq:fss_n_lead}
\end{equation}
where $d$ and $\omega^{\prime}$ are additional fitting parameters.
The obtained critical exponents $\eta_{\psi}$
for various choices of $L_{\text{min}}$ are summarized 
in Tables~\ref{tbl:n_without} and \ref{tbl:n_with},
and also in Fig.~\ref{fig:n-Lmin}.
We find that the stable solutions with $\omega^{\prime}>0$ 
can not be obtained for small $L_{\text{min}}$. Moreover, 
the estimated $\omega^{\prime}$ tends to increase
for larger $L_{\text{min}}$ (see Table~\ref{tbl:n_with}) and 
it is not possible to reach a converged value of 
$\omega^{\prime}$ within the given statistical accuracy and the system sizes available. 
Nevertheless, we confirm 
that the estimated values of $\eta_{\psi}$ 
with the correction term are instead converged and fully consistent 
with those obtained without
the correction term for large enough $L_{\text{min}}$,
as shown in Fig.~\ref{fig:n-Lmin}. 
Therefore, we only show in Sec.~\ref{subsec:fss_n} the results of the data collapse analysis 
without the leading correction term. 

\begin{table}[ht]
\caption{\label{tbl:m_without}%
 Results of the critical points $U_{\text{c}}/t$ and
 the critical exponents, $\nu$ and $\beta$,
 obtained from collapsing data of the staggered magnetization, $m_{\text{s}}(u, L)$,
 without the leading correction term, i.e., $c=0$ in Eq.~(\ref{eq:fss_m}), 
 for the honeycomb lattice model (upper rows)
 and the $\pi$-flux model  (lower rows).
 $L_{\text{min}}$ refers to the smallest $L$ used in the data collapse. 
 The maximum $L$ is 36 for the honeycomb lattice model 
 and 40 for the $\pi$-flux model. 
}
 \begin{ruledtabular}
 \begin{tabular}{l l l l}
  $L_{\text{min}}$ & $U_{\text{c}}/t$ & $\nu$ & $\beta$ \\
  \hline
   6       & 3.782(3)         & 1.025(4)  & 0.881(4)  \\
   9       & 3.785(5)         & 1.037(5)  & 0.886(6)  \\
  12       & 3.800(6)         & 1.041(5)  & 0.876(7)  \\
  15       & 3.820(7)         & 1.038(7)  & 0.856(10) \\
  18       & 3.833(9)         & 1.040(11) & 0.841(14) \\
  \\
   8       & 5.415(10)        & 0.999(9)  & 0.873(9)  \\
  12       & 5.418(14)        & 1.018(10) & 0.886(12) \\
  16       & 5.455(21)        & 1.011(11) & 0.861(17) \\
  20       & 5.509(24)        & 1.025(12) & 0.838(20) \\
  24       & 5.511(37)        & 1.053(22) & 0.855(36) \\
 \end{tabular}
 \end{ruledtabular}
\end{table}

\begin{table}[ht]
\caption{\label{tbl:m_with}%
 Same as Table.~\ref{tbl:m_without},
 but with the leading correction term $c L^{-\omega}$ ($c \neq 0$)
 in Eq.~(\ref{eq:fss_m}).
}

 \begin{ruledtabular}
 \begin{tabular}{l l l l l }
  $L_{\text{min}}$   & $U_{\text{c}}/t$ & $\nu$     & $\beta$  & $\omega$ \\
  \hline
   6 & 3.843(8)  & 1.005(5)  & 0.74(2)  & 0.55(4)  \\
   9 & 3.858(9)  & 1.012(5)  & 0.74(2)  & 0.78(5)  \\
  12 & 3.856(10) & 1.020(7)  & 0.75(2)  & 0.91(5)  \\
  15 & 3.853(10) & 1.021(8)  & 0.75(2)  & 0.89(6)  \\
  18 & 3.849(10) & 1.028(10) & 0.76(2)  & 0.82(12) \\
  \\
   8 & 5.423(38) & 0.998(10) & 0.86(5)  & 0.17(35) \\
  12 & 5.534(41) & 1.007(10) & 0.76(5)  & 0.94(25) \\
  16 & 5.557(31) & 1.008(11) & 0.74(3)  & 1.02(13) \\
  20 & 5.546(27) & 1.021(11) & 0.76(3)  & 0.85(24) \\
  24 & 5.537(35) & 1.050(19) & 0.78(4)  & 0.83(17) \\
 \end{tabular}
 \end{ruledtabular}
\end{table}

\begin{table}[ht]
\caption{\label{tbl:n_without}%
 Results of the critical exponent $\eta_{\psi}$
 obtained from collapsing data of 
 the jump in the energy resolved momentum distribution function, $\Delta n(u,L)$,
 without the leading correction term, i.e., $d=0$ in Eq.~(\ref{eq:fss_n_lead}), 
 for the honeycomb lattice model (left rows)
 and the $\pi$-flux model  (right rows).
 $L_{\text{min}}$ refers to the smallest $L$ used in the data collapse. 
 The maximum $L$ is 36 for the honeycomb lattice model 
 and 40 for the $\pi$-flux model. 
 }
 \begin{ruledtabular}
 \begin{tabular}{l l l l}
  \multicolumn{2}{l}{honeycomb lattice} & \multicolumn{2}{l}{$\pi$-flux model} \\
  $L_{\text{min}}$ & $\eta_{\psi}$      & $L_{\text{min}}$ & $\eta_{\psi}$     \\
  \hline
   6 & 0.17(2) &  8 & 0.19(2) \\
   9 & 0.18(2) & 12 & 0.21(2) \\
  12 & 0.19(2) & 16 & 0.22(2) \\
  15 & 0.20(2) & 20 & 0.23(2) \\
  18 & 0.20(2) & 24 & 0.24(2) \\
  \end{tabular}
 \end{ruledtabular}
\end{table}

\begin{table}[ht]
\caption{\label{tbl:n_with}%
 Same as Table.~\ref{tbl:n_without},
 but with the leading correction term $d L^{-\omega^{\prime}}$ ($d \neq 0$)
 in Eq.~(\ref{eq:fss_n_lead}).
 }
 \begin{ruledtabular}
 \begin{tabular}{l l l  l l l}
  \multicolumn{3}{l}{honeycomb lattice}  & 
  \multicolumn{3}{l}{$\pi$-flux model }  \\
  $L_{\text{min}}$ & $\eta_{\psi}$ & $\omega^{\prime}$  & 
  $L_{\text{min}}$ & $\eta_{\psi}$ & $\omega^{\prime}$  \\
  \hline
  12 & 0.25(5) & 1.3(6) & 16 & 0.41(10)& 0.5(6) \\
  15 & 0.23(3) & 1.6(3) & 20 & 0.33(8) & 1.0(5) \\
  18 & 0.23(3) & 1.7(3) & 24 & 0.31(4) & 1.2(3) \\
 \end{tabular}
\end{ruledtabular}
\end{table}

\bibliography{mit}

\begin{thebibliography}{70}
\bibitem [{squ()}]{square}%
  \BibitemOpen
  \href@noop {} {}\bibinfo {note} {For instance, in the square lattice with the
  nearest-neighbor hopping, the perfectly nested Fermi surface induces an
  insulating phase once non-zero $U$ is introduced.}\BibitemShut {Stop}%
\bibitem [{sin()}]{singularity}%
  \BibitemOpen
  \href@noop {} {}\bibinfo {note} {The singularity in the momentum distribution
  function implies a long-distance behavior of the density matrix $\langle
  c_{is}^\dag c_{js}\rangle \sim Z/|\bm{r}_i-\bm{r}_j|^2$, the same kind of
  behavior implied for a single-particle Green's function $G(\bm{k},\omega)$
  with the residue $Z$ of a quasiparticle pole, obtained after a careful
  integration in $\bm{k}$ and $\omega$. Here $\bm{r}_i$ is the location of site
  $i$.}\BibitemShut {Stop}%
\bibitem [{Not()}]{Note_eta}%
  \BibitemOpen
  \href@noop {} {}\bibinfo {note} {By fitting the quasiparticle weight $Z$ in a
  form given in Eq.~(\ref{eq:z_eta}), we have found that $\eta_{Z}=0.44\pm0.16$
  for the honeycomb lattice model and $\eta_{Z}=0.59\pm0.17$ for the $\pi$-flux
  model. Noticing that $\eta_Z=\nu\eta_\psi$, these values are almost two times
  larger than the ones estimated by the data collapse fits in
  Sec.~\ref{sec:fss} (also see in Sec.~\ref{sec:conclusion}). This difference
  is mostly due to the non trivial extrapolation procedure for the estimation
  of $Z$: $Z$ is estimated by extrapolating $Z_L$ in the thermodynamic limit
  with a polynomial function of $1/L$, where $Z_L$ is first obtained by
  extrapolating $n(\varepsilon_{\bm k})$ to the Fermi level, thus involving two
  extrapolations. Therefore, we rely more on the values obtained by the
  finite-size scaling analysis in Sec.~\ref{sec:fss}, where i) no extrapolation
  procedure is involved and ii) all data points are used not only for
  $U<U_{\text{c}}$ but also for $U>U_{\text{c}}$.}\BibitemShut {Stop}%
\bibitem [{not()}]{note_nq}%
  \BibitemOpen
  \href@noop {} {}\bibinfo {note} {For example, in the
  exact infinite dimension calculation, which is well known to be equivalent to
  the Gutzwiller approximation, it can be shown that $g_{\text{NN}}= (4 \bar
  d)^2$~\cite{gebhard}.}\BibitemShut {Stop}%
\end{thebibliography}
\end{document}